\documentclass[
 floatfix, 
 reprint,
 amsmath,amssymb,
 aps,
]{revtex4-2}
\usepackage{algorithm}    
\usepackage{algpseudocode} 
\usepackage{graphicx}
\usepackage{dcolumn}
\usepackage{bm}

\usepackage{lineno}
\usepackage{ragged2e}

\newcommand{\Surviving}{\mathcal{S}}
\newcommand{\Extinct}{\mathcal{E}}
\newcommand{\Invading}{\mathcal{I}}

\usepackage{xcolor}
\usepackage{graphicx}

\newcommand{\Akshit}[1]{\textcolor{black}{#1}}

\begin{document}

\title{A theory of ecological invasions and its implications for eco-evolutionary dynamics}

\author{Zhijie Feng$^1$}
\author{Emmy Blumenthal$^{1,2}$}
\author{Pankaj Mehta$^{1,3}$}
\email{pankajm@bu.edu}
\author{Akshit Goyal$^4$}
\email{akshitg@icts.res.in}

\affiliation{$^1$Department of Physics, Boston University, Boston, Massachusetts 02215, USA}
\affiliation{$^2$Joseph Henry Laboratories of Physics, Princeton University, Princeton, NJ 08544, USA}
\affiliation{$^3$Faculty of Computing and Data Sciences, Boston University, Boston, Massachusetts 02215, USA}
\affiliation{$^4$International Centre for Theoretical Sciences, Tata Institute of Fundamental Research, Bengaluru 560089, India}

\date{\today}

\begin{abstract}
Predicting the outcomes of species invasions is a central goal of ecology, a task made especially challenging due to ecological feedbacks. To address this, we develop a general theory of ecological invasions applicable to a wide variety of ecological models: including Lotka-Volterra models, consumer resource models, and models with cross feeding. Importantly, our framework remains valid even when invading evolved (non-random) communities and accounts for invasion-driven species extinctions. We derive analytical expressions relating invasion fitness to invader abundance, shifts in the community, and extinction probabilities. These results can be understood through a new quantity we term ``dressed invasion fitness'', which augments the traditional notion of invasion fitness by incorporating ecological feedbacks. We apply our theory to analyze short-term evolutionary dynamics through a series of invasions by mutants whose traits are correlated with an existing parent. We demonstrate that, generically, mutants and parents can coexist, often by driving the extinction of low-abundance species. We validate theoretical predictions against experimental datasets spanning ecosystems from plants to microbial protists. Our work highlights the central role of ecological feedbacks in shaping community responses to invasions and mutations, suggesting that parent-mutant coexistence is widespread in eco-evolutionary dynamics.
\end{abstract}

\maketitle

% \linenumbers
\section{Introduction}

Invasion by a new species is a fundamental feature of ecosystems ranging from microbiomes to rainforests \cite{tilman2004niche,melbourne2007invasion,hoffmann2016biological}. Ecological invasions also offer a powerful lens through which to view ecological and evolutionary processes \cite{sakai2001population,sax2007ecological}. These include practical engineering tasks like ecosystem management  \cite{strayer2012eight} and designing probiotics \cite{goyal2018multiple}, but also more general ecological questions regarding community function and resilience \cite{richardson2008fifty, lockwood2013invasion}. More abstractly, a single step of evolution can also be viewed through the lens of invasions by viewing new mutations as invasions by highly related species \cite{mcenany2024predicting, mahadevan2024continual, good2023eco}. 

Despite the importance of ecological invasions, our theoretical understanding remains limited \cite{vellend2007effects,shea2002community,facon2006general,maron2004rapid}. This gap in our knowledge is even more pronounced when the communities of interest are diverse, evolved, and/or engage in complex behaviors such as cross-feeding \cite{goyal2018diversity}.  Invasion outcomes are mediated by a variety of factors including invader characteristics, community interactions, and environmental conditions \cite{hu2025collective, sax2007ecological, case1990invasion, kurkjian2021impact}.  A further complication that arises when thinking about invasions \Akshit{is the presence of ecological feedbacks, i.e., the fact that invaders fundamentally alter the ecosystem they invade} \cite{ives2007stability,phillips2006invasive}.

A prominent example of this is invasion-induced extinction of species originally present in the community \cite{doherty2016invasive,bellard2016alien,davis2003biotic}. Dealing with extinctions presents an especially challenging technical problem. It is hard to {\it a priori}  predict whether an existing species will survive or go extinct after an invasion.  This is especially true in context of evolution where the invader is highly correlated with an existing species in the community \cite{good2023eco,mcenany2024predicting}. 

\begin{figure*}[t]
\centering
\includegraphics[width=\textwidth]{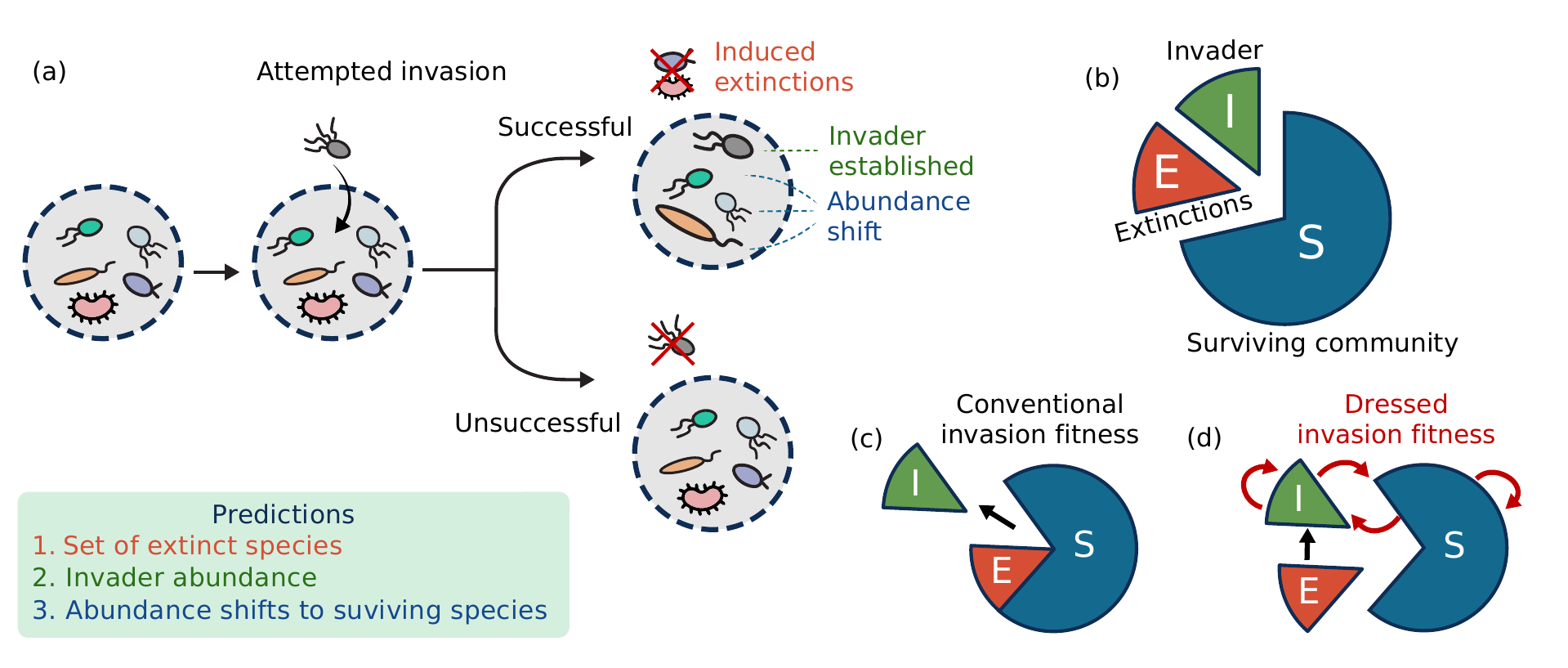}
 \caption{\textbf{Predicting invasion outcomes in diverse ecological communities.} (a) Cartoon illustrating possible invasion outcomes. If an attempted invasion is unsuccessful, the community remains unchanged; if successful, the invader establishes, causing extinctions and abundance shifts among survivors. Our theory predicts all these three outcomes: the set of species that go extinct, the invader abundance, and the abundance shifts among surviving species. (b) We show that prediction requires partitioning a community into extinct and surviving species post-invasion (Eq \eqref{eq:4 block}). (c) We introduce ``dressed invasion fitness'' (Eq \eqref{eq:dressedIF}), which incorporates ecological feedbacks and show it predicts invasion outcomes better than conventional invasion fitness, which only considers direct community interactions. }
\label{fig:invasion_fig_doodles}
\end{figure*}

To overcome these challenges, here we develop a general framework for modeling ecological invasions with species extinctions.  Our framework allows us to predict invader abundances, which species go extinct, and changes in community abundances by solving a set of self-consistency equations relating these quantities to each other (see Fig. \ref{fig:invasion_fig_doodles}(a)). The framework takes advantage of the fact that, for diverse ecosystems, we can view invasions as small perturbations. This allows us to understand invasion dynamics using linear-response theory \cite{Goyal_2025}, now augmented with the ability to account for invasion-induced extinctions.

A key advantage of our approach is its applicability to evolved communities and invaders with traits closely related to existing species in the ecosystem, which we henceforth refer to as correlated invaders. Such invaders arise during evolution, where mutants are highly-correlated with parents. We demonstrate that parent-mutant coexistence is mediated by interactions with the \emph{entire community} through global eco-evolutionary feedbacks. A direct consequence of this is that for diverse communities, parents and mutants can often coexist despite being more phenotypically similar to each other than other community members. 

Importantly, our theory is compatible with almost any niche-based ecological model where the species and resources reach a steady state (e..g, Lotka-Volterra and consumer resource models \cite{Blumenthal_2023}, models with cross-feeding \cite{goldford2018emergent, Pacheco_2019, muscarella2020species, marsland2019available}, and models with more complex non-linearities \cite{cui2024houches}).  Our theory also makes direct connection with experiments, highlighting its potential for connecting theoretical predictions with empirical observations.

\section{Theory of ecological invasions}

Our goal is to describe how ecosystems respond to invasions by new species. We assume that the ecosystem being invaded is at steady state (i.e. resource and species dynamics are not changing in time) and ask about the new steady state reached by the system after invasion.  As illustrated in the cartoon in Fig.  \ref{fig:invasion_fig_doodles} (a), there are two possible outcomes post-invasion: if the invasion fails, the system eventually recovers to the same ecological steady state as before;  if the invasion succeeds, the invader proliferates and establishes itself within the system, leading to changes in community structure and the abundances of surviving species and resources. 
 
 In this work, we introduce a general framework to predict the new ecological steady state after an invasion, making use of the knowledge of the pre-invasion community and their interactions with the invader.  The outcomes of successful invasions can be categorized by three experimentally measurable quantities: (1) the identity of species that go extinct, (2) the shifts in abundance among surviving species, and (3) the abundance of the invader itself. When the pre-invasion ecosystem \Akshit{is sufficiently diverse (see Fig.\ref{fig:error plot}) and at steady-state, the introduction of invaders can be viewed as adding a small initial perturbation to the existing ecosystem}. Our framework exploits this observation to predict invasion outcomes systematically using a linear-response theory for ecological invasions, \Akshit{which works even though invaders that start at low abundances can eventually grow to large abundances. Our assumption of linear response is justified whenever the effective ecological interactions in pre- and post-invasion communities are approximately similar (see Appendix~\ref{app:general framework} for precise definition). For many canonical ecological models (e.g., the Lotka-Volterra and MacArthur consumer resource models), we can show that this assumption is exactly true, not just approximate (Appendix~\ref{app:general framework} and \ref{app: derive CRM response}). For a wide variety of other models, we find that this assumption is often approximately valid (see Fig.~\ref{fig:all models agree} and Fig.~\ref{fig:error plot}). As a result, our framework can successfully predict invasion outcomes even in experimental data from low-diversity communities with $<10\%$ error (see Section V).}
 
\subsection{Predicting invasion outcomes}
Our framework is broadly applicable to a wide range of ecological models but in the main text we focus on the generalized Lotka-Volterra model (GLV). The dynamics of the GLV model are given by 
\begin{align}
    \frac{dX_i}{dt} \equiv X_i g_i(\vec{X}) =X_i\left( r_i- X_i - \sum_{j\neq i} A_{ij}X_j \right),
\label{eq:GLV-main}
\end{align}
where $i$ labels species from 1 to $S$, $X_i$ is the abundance of species $i$ and $g_i(\vec{X})$ is its net per capita growth rate in the presence of other species. $g_i(\vec{X})$ is a linear function of its naive growth rate $r_i$ and the abundances of all species $X_j$. The interaction coefficients $A_{ij}$ characterize how the presence of species $j$ affects the growth rate of species $i$. 
Throughout this study we assume that the dynamics converge to a steady state, \Akshit{but it need not be unique. Our framework works equally well for ecosystems with multiple stable steady states (see Fig.~\ref{fig:multistability_combined})}. Denote the pre-invasion abundances of the species in the community by ${X}^{\text{\text{old}}}_j$.  We then ask what happens when we invade such an ecosystem by a new species, denoted by the distinguished index $0$.  Within GLV, the invader is fully characterized by specifying its interaction coefficients $A_{0j}$ and $A_{j0}$ and its growth rate $r_0$. 

Our goal is to predict whether the invasion will be successful, and if successful, the set of species that survive (denoted by the set $ \Surviving$), the set of species that go extinct (denoted by $ \Extinct$), the invader abundance $X_0$, and the abundances of surviving species post invasion $\vec{X}_\Surviving^{\text{new}}={X}^{\text{\text{old}}}_\Surviving+\delta \vec{X}_\Surviving$ for $j \in \Surviving$. Note that the quantity $\delta \vec{X}_\Surviving$ characterizes the shift in species abundances pre- and post-invasion.

To achieve our goal, we will write a set of self-consistency equations for our quantities of interest. Assume we know the outcome of the invasion, and in particular which species survive, $\Surviving$, and which species go extinct, $\Extinct$. Then we can split the pre-invasion interaction matrix of the community into four blocks, \begin{align}
 A^{\text{\text{old}}}=\begin{pmatrix}
A_{\Surviving\Surviving}&A_{\Surviving\Extinct}\\
A_{\Extinct\Surviving}&A_{\Extinct\Extinct}
\end{pmatrix}.
\label{eq:4 block}
\end{align}
We also partition the pre-invasion species abundances and $\vec{r}_j$ into surviving and extinct species: $\vec{X}^{\text{old}}_{\Surviving}$, $\vec{X}^{\text{old}}_{\Extinct}$, and $\vec{r}_\Surviving, \vec{r}_\Extinct$.  Having defined these partitions, the post-invasion steady-state condition becomes \begin{equation}
   g_\Surviving(\vec{X}^{\text{new}})= \vec{r_\Surviving}-A_{\Surviving\Surviving}\vec{X}^{\text{new}}_\Surviving-A_{\Surviving0}{X}_0=0.
    \label{eq: surviving equality}
\end{equation}
We also know that extinct species have a negative invasion fitness, namely the will have a negative growth rate if they are re-introduced
into the ecosystem
\begin{equation}
    g_\Extinct(\vec{X}^{\text{new}})=\vec{r_\Extinct}-A_{\Extinct\Surviving}\vec{X}_\Surviving^{\text{new}}-A_{\Extinct0}{X}_0 <  0.
    \label{eq: extinct inequality}
\end{equation}

Our key insight is to note that even though the abundances of extinct species $X_\Extinct$ change abruptly to zero post-invasion, the post-invasion growth rates of extinct species $g_\Extinct(\vec{X}^{\text{new}})$ change more smoothly from positive to negative and can hence still be calculated perturbatively. 

Drawing on these observations, in Appendix \ref{app: derive invader abundance}  we show the invader abundance can be expressed in terms of the quantities introduced above as
\begin{align}
     X_0&=\frac{ r_0 - A_{0\Surviving}\vec{X}_{\Surviving}^{\text{old}}-A_{0\Surviving}A^{-1}_{\Surviving \Surviving} A_{\Surviving \Extinct} \vec{X}_{\Extinct}^{\text{old}}}{A_{0 0} - A_{0\Surviving}A^{-1}_{\Surviving \Surviving}A_{\Surviving 0}},
     \label{eq:invader abundance}
\end{align}
and that the change in the abundances of surviving species is given by
\begin{equation}
    \delta \vec{X}_{\Surviving}=A_{\Surviving \Surviving}^{-1} ( -A_{\Surviving 0} {X}_{0}+A_{\Surviving \Extinct} \vec{X}^{\text{old}}_{\Extinct}).
     \label{eq:community shift}
\end{equation}
The quantity in the bracket of  Eq.~\eqref{eq:community shift} can be understood as an effective environmental perturbation that incorporates numerous ecological feedbacks resulting from invasion. This quantity can be decomposed into two parts: a term proportional to the invader abundance ${X}_{0}$ that captures the effects of feedbacks due to the invader and a term proportional to $\vec{X}^{\text{old}}_{\Extinct}$ that captures changes in the environment due to invasion-induced extinctions. Notice that this environmental perturbation is converted to shifts in abundances by the \emph{inverse} of the interaction matrix of surviving species $A_{\Surviving \Surviving}^{-1}$. This is a global quantity that depends on the properties of \emph{all the surviving species $\Surviving$}, highlighting that the outcomes of invasion depend intricately on the structure of all community members in an extremely complicated and non-local manner.

\Akshit{Note that we do \emph{not} assume that invaders always cause a weak effect on a community: we explicitly include the strong effects of invaders on communities over the invasion process. Invaders exert a weak effect on the community \emph{initially}, i.e., they invade at low abundance, as is common assumption across ecological literature \cite{arnoldi2022invasions}. However, as invaders grow their effect becomes strong and can cause large changes to communities, such as extinctions of other resident species. These effects appear explicitly in Eqs. (\ref{eq:invader abundance})--(\ref{eq:community shift}), as surviving species $\mathcal{S}$ behave differently from extinct ones $\mathcal{E}$.} To close this set of equations, we impose self-consistency. Namely, we require that surviving species $\Surviving$ have positive abundances
\begin{align}
\vec{X}_\Surviving^{\text{new}} =\delta \vec{X}_\Surviving+ \vec{X}_\Surviving^{\text{old}}> 0,
\label{eq:surviving-SCE}
\end{align}
and that the invasion fitness of the extinct species are negative (Eq. \eqref{eq: extinct inequality}). 

Collectively, Eqs.~\eqref{eq: extinct inequality}-\eqref{eq:surviving-SCE} define a set of self-consistent equations (a mix of equalities and inequalities) that must be satisfied by the post-invasion abundances. \Akshit{Fully solving these equations requires an exhaustive combinatorial scan that is computationally expensive. Instead, we use a simpler iterative approach (details in Appendix I Algorithm.~\ref{alg:general_perturbation}) which is common in constrained optimization because of its efficiency and  accuracy ($\sim 99\%$ in our case).}
We will later show that these self-consistency equations have an natural ecological interpretation in terms of a new ecological quantity we call the ``dressed invasion fitness'' that generalizes the notion of invasion fitness to account for community-mediated ecological feedbacks.

\subsection{Tests of model predictions}
\begin{figure*}
\centering
\includegraphics[width=1\textwidth]{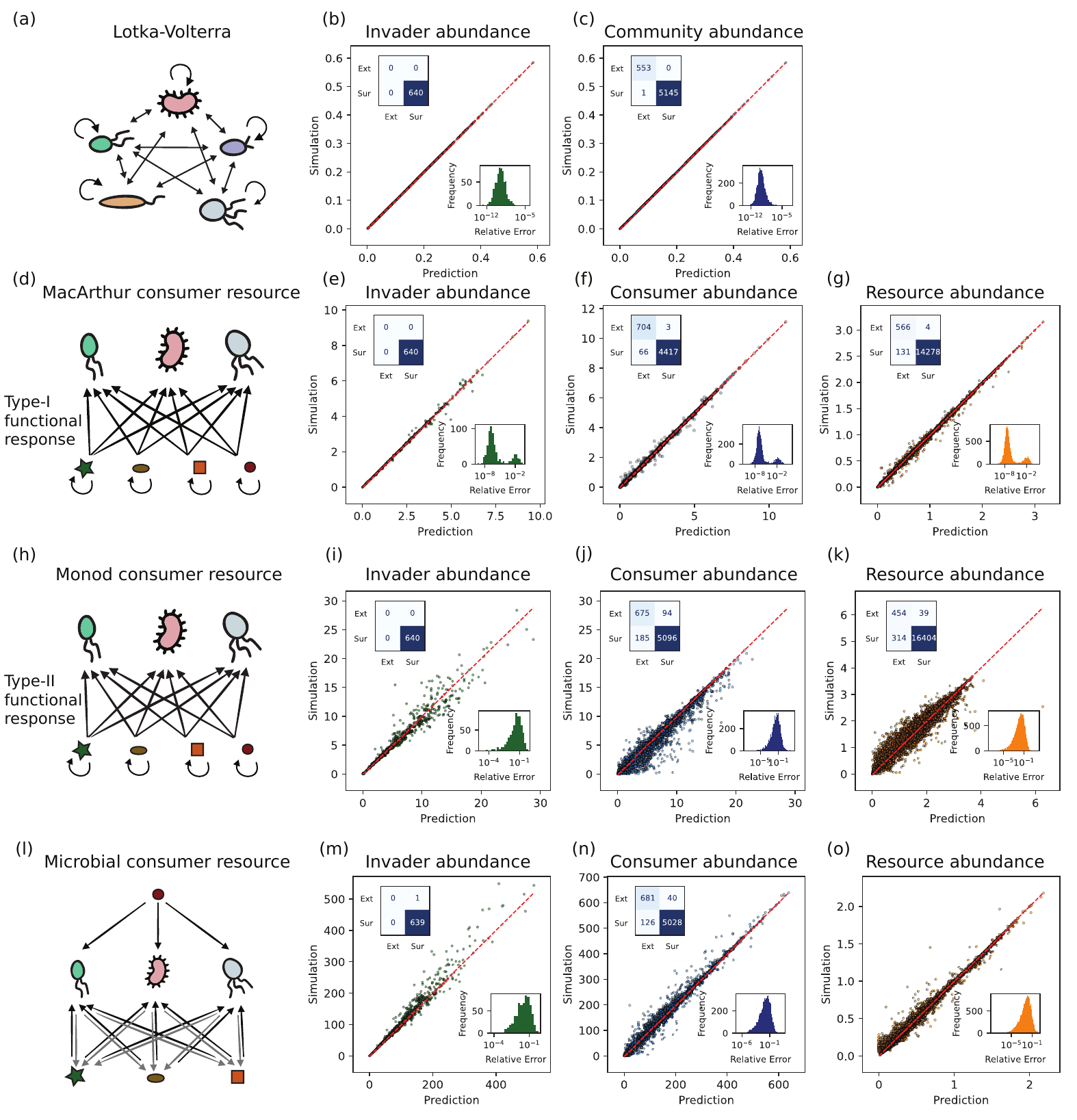}
 \caption{\textbf{Theory makes accurate predictions across diverse ecological models.} \Akshit{Comparison between simulations theoretical predictions for (a--c) Lotka-Volterra models,  (d--g) MacArthur consumer resource models with Type I functional response, (h--k) Monod consumer resource with Type II functional response (Monod dependence of growth rate on resource abundances), and (l--o) microbial consumer resource models with cross-feeding. Panels display simulated versus predicted abundances for invaders, community residents and resources respectively. Each data point indicates one species/resource. Insets for each panel show (1) matrices indicating qualitative prediction outcomes (extinct or surviving; rows indicate simulations, columns predictions); and (2) histograms of relative errors.}
 }
\label{fig:all models agree}
\end{figure*}

In the last section, we derived a set of self-consistency equations describing ecological invasions for the Lotka-Volterra model. In Appendix \ref{app:nonlinear-model}, we show that self-consistency equations analogous to Eqs.~\eqref{eq: extinct inequality}-\eqref{eq:surviving-SCE} can be derived for a large class of niche-based ecological models, including most variants of consumer resource models. In this more general setting, the pairwise interaction matrix $A$ is replaced by an inverse of an effective interaction matrix that captures how a perturbation in the abundance of one species affects the abundances of other species. \Akshit{In addition, for consumer resource models, the self-consistency equations for species must be augmented by additional equations describing the response of resources to  invasion. Nonetheless, the qualitative logic, derivation, and structure of the self-consistency equations remains the same.}

To demonstrate the generality of our framework, we compare our theoretical predictions to numerical simulations for \Akshit{four} different classes of niche-based models in theoretical ecology: (i) the Generalized Lotka-Volterra model, \Akshit{(ii) the MacArthur consumer resource model with type I functional response; the (iii) the Monod consumer resource model with type II functional response, and (iv) the Microbial consumer resource model with cross-feeding (see Fig. \ref{fig:all models agree}). }

\emph{Lotka-Volterra models.} We simulated $640$ communities with $10$ species and dynamics given by the generalized Lotka-Volterra equations in Eq.~\eqref{eq:GLV-main}. We focused on ecosystems with symmetric interactions where the interaction coefficients $A_{ij}$  were drawn from a random normal distribution (see Appendix \ref{app: LV numerical} for details of parameters). After running the dynamics until steady state, on average $8$ of the $10$ original species survived in our simulations. We then introduced an invader into the ecosystem and compared theoretical predictions to numerical simulations. We focused on three quantities: (a) predicting whether the invader would invade successfully,  (b) the post-invasion invader abundance for successful invasions, and (c) the resulting shift in the species abundances (the community shift). As can be seen in Fig. \ref{fig:all models agree} (b)-(c), the theory and simulations agree remarkably well.

\emph{MacArthur consumer resource models.} Next, we applied our theoretical framework using the MacArthur consumer resource model (MCRM) \cite{chesson1990macarthur, macarthur1967limiting}. Like the Lotka-Volterra model, the MCRM is a foundational model of niche theory.  In MCRMs, species with abundances $N_i$ deplete resources with abundances $R_\alpha$. Species $i$ have preferences $C_{i \alpha}$  for resource $\alpha$. We further assume that in the absence of consumers, the resource dynamics can be described by a Lotka-Volterra equation. Within this model, all interactions between species are mediated by competition for resources, with dynamics given by
\begin{align}
\frac{dN_i}{dt}&=N_i\left(\sum_\alpha C_{i\alpha}R_\alpha-m_i\right)\nonumber \\
\frac{dR_\alpha}{dt}&=R_\alpha\left(K_\alpha - \sum_{\beta} Q_{\alpha \beta} R_\beta\right)-\sum_j C_{j\alpha}N_jR_\alpha,
\label{eq: general MCRM}
\end{align}
where $m_i$ is the mortality rate of species $i$, $K_\alpha$ is the intrinsic growth rate of resource $\alpha$, and $Q_{\alpha \beta}$ is the resource-resource interaction matrix. Fig. \ref{fig:all models agree}(d) presents a schematic representation of a typical MacArthur consumer-resource model. 

To test these theoretical predictions, we simulated $640$ communities using Eq.~\eqref{eq: general MCRM}, with parameters drawn randomly (see Appendix \ref{app: CRM numerical} for details). We initialized each community with $60$ consumers and 30 resources and ran the dynamics to steady state. After this initial phase, we introduced an invader and simulated the post-invasion ecosystem until it reached a steady state. We then compared our theoretical predictions for invasion success, invader abundance, and the shifts in species and resource abundances with the results of our simulation (see Fig. \ref{fig:all models agree} (e)-(g)). As in the Lotka-Volterra model, our theory for the MCRM recapitulates the simulation results. \Akshit{The error distributions for this model (insets of Figs.~\ref{fig:all models agree}(e--g)) have two modes: the left peak ($\sim10^{-6}\%$) is due to numerical errors, while the right peak ($\sim 1\%$) is due to occasional errors in our iterative approach to solve the self-consistency equations (a non-iterative approach removes it; see Fig.~\ref{fig:MCRM_work_with_info}).}

\Akshit{\emph{Monod model and Microbial consumer resource model with cross-feeding.} Finally, to test our theory in more complex ecological settings, we applied our framework to two additional models: a consumer resource model with Monod or type II functional response, and the Microbial consumer resource model (MiCRM) \cite{marsland2019available, MehtaCrossfeeding}. The Monod model extends the MacArthur model to include a nonlinear dependence of species' per capita growth rates on resource abundances (Appendix~\ref{app:nonlinear-model}). The \mbox{MiCRM} extends traditional consumer resource models by explicitly modeling cross-feeding by microbial species --- where metabolic byproducts produced by one species are consumed by others. As a result of these extensions, the effective interaction matrix $A^{\text{eff}}$ in these models depend on community state and are not fixed. In these cases, our theory assumes that the post-invasion $A^{\text{eff}}$ is still approximately the same as the one pre-invasion. We can thus use the same steps as before to derive self-consistency equations for invader abundance and community abundance shifts (details of approximations and derivation for both models in Appendix \ref{app:nonlinear-model}).  Fig. \ref{fig:all models agree}(i--o) shows that despite approximations, there is remarkably good agreement between theoretical predictions and simulations (simulation details in Appendix \ref{app: MiCRM numerical}). These results suggest that our theory applies to a variety of ecological models and successfully predicts all key properties of invasion outcomes.}

\section{Dressed invasion fitness governs outcomes}

\begin{figure*}
\centering
\includegraphics[width=1\textwidth]{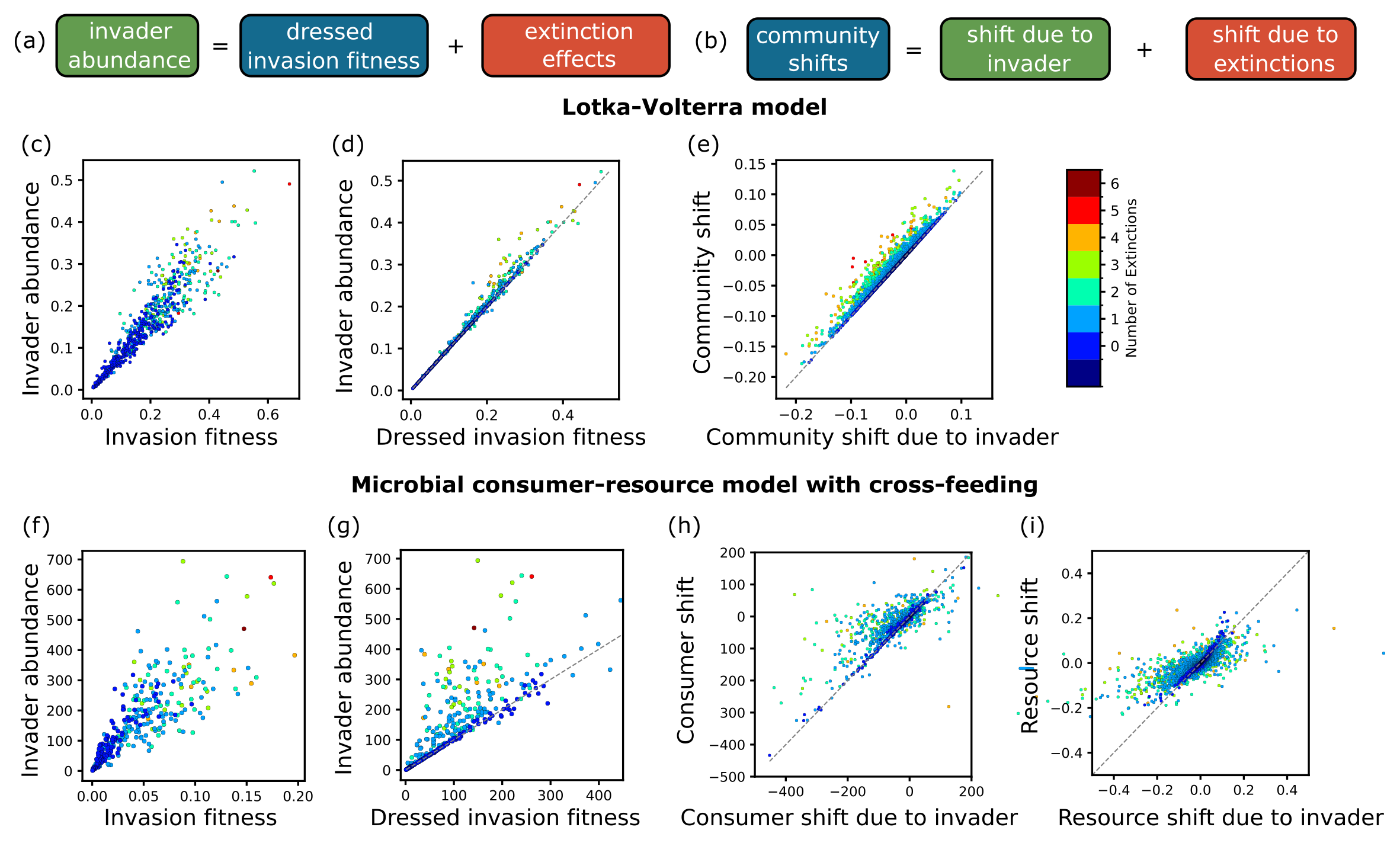}
 \caption{\textbf{Dressed invasion fitness and extinction-induced feedbacks govern invasion outcomes.} Our theory shows that both (a) invader abundance and (b) shifts in community abundances can be decomposed into two parts: a direct contribution the invader --- through the dressed invasion fitness --- and an indirect contribution due to extinctions (Eqs (11)). (c--i) Both Lotka-Volterra and Microbial consumer resource model simulations that (c--d) invader abundance is generally correlated with naive invasion fitness, but (d) and (g) show that dressed invasion fitness --- which correctly incorporates ecological feedbacks --- is a much better predictor of invader abundance.  Similarly (e) and (h--i) show that our theory can also generally predict shifts in community abundances through the invader's direct effect (dashed line and blue dots) in the absence of extinctions. To predict the effect of invaders that introduce more extinctions --- which show deviations from direct effect-based predictions (colored dots) ---  we need to account for indirect effects as well (Eq. \eqref{eq:community shift}). \Akshit{Fig.~\ref{fig:SI shift and dressed invasion fitness} shows analogous plots for the the MacArthur and Monod consumer resource models}. }
\label{fig:extinction effect}
\end{figure*}

We now provide a simple ecological interpretation of the outcomes of our theory in terms of a new quantity we call the ``dressed invasion fitness''. As the name suggests, the dressed invasion fitness extends the classic concept of invasion fitness, the growth rate of an invader when it is initially introduced into a resident ecosystem, by incorporating the effects of ecological feedbacks. 

Consider an invading an ecosystem governed by Eq.~\eqref{eq:GLV-main} by a new species $0$ with naive growth rate $r_0$ and interaction matrix $A_{0j}$. The invasion fitness $g_0^{\text{naive}}$ of species $0$ is given by
\begin{align}
    g_0^{\text{naive}} &= r_0 - \sum_j A_{0j}x^{\text{old}}_j,
    \label{eq:g0naive}
\end{align}
where $x^{\text{old}}_j$ is the pre-invasion abundance of species $j$. A negative invasion fitness ($g_0^{\text{naive}}< 0$) means that an invader cannot successfully invade an ecosystem. However, knowing an invader has a positive invasion fitness is not enough to predict its final abundance. This is because invasion fitness depends only on the pre-invasion state of the ecosystem and therefore is agnostic to the subsequent ecological dynamics resulting from feedback and extinction.

Nonetheless, ecological intuition suggests that there should be a relationship between the final invader abundance $X_0$  and the invasion fitness $g_0^{\text{naive}}$ since species with a high initial growth rate are more likely to end up at high abundance once the ecosystem reaches its new post-invasion steady state. We find that this is indeed the case. As before, if we denote the set of species that survive after a successful invasion by $\Surviving$ and  the set of species which go extinct by $\Extinct$,  a straight forward calculation shows that it is possible to rewrite Eq.~\eqref{eq:invader abundance} for $X_0$ as
\begin{align}
X_0&=g^{\text{dressed}}_0+ \vec{c}_\Extinct \cdot \vec{X}_{\Extinct}^{\text{old}},
\label{eq:x0interp}
\end{align}
where we have defined the ``dressed invasion fitness''
\begin{align}
    g^{\text{dressed}}_0=\frac{g_0^{\text{naive}}}{A_{00} - A_{0\Surviving}A^{-1}_{\Surviving \Surviving}A_{\Surviving 0}},
    \label{eq:dressedIF}
\end{align}
 $\vec{X}_{\Extinct}$ is a vector of size $|\Extinct|$ consisting of the pre-invasion abundances of species that go extinct during the invasion, and
\begin{align}
	\vec{c}_\Extinct=\frac{A_{0\Extinct}+A_{0\Surviving}A^{-1}_{\Surviving \Surviving} A_{\Surviving \Extinct}}{A_{00} - A_{0\Surviving}A^{-1}_{\Surviving \Surviving}A_{\Surviving 0}}.
\end{align}
Eq.~\eqref{eq:x0interp} shows that the invader abundance post-invasion can be decomposed into a sum of two terms: (1) a term proportional to the invasion fitness, which we call the dressed invasion fitness $g^{\text{dressed}}_0$; and (2) a term due to species extinctions $ \vec{c}_\Extinct \cdot \vec{X}_{\Extinct}^{\text{old}}$ (Fig. ~\ref{fig:extinction effect}a). \Akshit{The dressed fitness, unlike naive fitness  $g_0^{\text{naive}}$, accounts for the propagation of feedback from invader to surviving community and back to invader (the second term in the denominator of Eq.~\eqref{eq:dressedIF}). While the naive fitness only measures the invader's initial growth rate at small abundance, the dressed invasion fitness accounts for the invader's effect on the community as its abundance gets larger. As an invader grows,  it starts to affect other community members, which affect it back. The dressed invasion fitness augments naive invasion fitness (numerator) with exactly this effect (denominator of Eq. (\ref{eq:dressedIF})).} Figs. \ref{fig:extinction effect} (c-d) show the invader abundance $X_0$ as a function of invasion fitness $g_0^{\text{naive}}$ and dressed invasion fitness $g_0^{\text{dressed}}$. As expected,  both these quantities are correlated with invader abundance. The dressed invasion fitness is much more predictive of the final invader abundance, especially in the absence of species extinctions (Fig. \ref{fig:extinction effect}d). This shows that the dressed invasion fitness which incorporates ecological feedbacks is really the ``right'' quantity to predict the fate of an invader.

Additionally, we can decompose the shift in the post-invasion abundances of surviving species  $\delta \vec{X}_{\Surviving}$ into two parts:  a term caused directly by the invader and a term resulting from extinctions (the terms on the right-hand side of Eq.~\eqref{eq:community shift} proportional to $X_0$ and $X_\Extinct$ respectively).  Fig. \ref{fig:extinction effect}(e) shows a plot of the observed community shift versus the first term of Eq.~\eqref{eq:community shift} that captures the portion of the community shift due to direct interaction with the invader. As expected, in the absence of extinctions, the predicted and observed values agree with deviations growing progressively larger as the number of extinctions increases.

Analogous interpretations exist for consumer-resource models, including \Akshit{the Monod model and  MCRM (see Appendix \ref{app: derive CRM response} -\ref{app: derive response MiCRM} and Figs.~\ref{fig:SI shift and dressed invasion fitness}, omitted here for brevity).} Once again all relevant quantities can be decomposed into a contribution due to the invader and a contribution due to extinctions. 
As can be seen in Figs. \ref{fig:extinction effect} (f--i), these decompositions continue to hold in consumer resource models despite the presence of more complex dynamics.

\section{Implications for eco-evolutionary dynamics}
\begin{figure*}[t]
\centering
\includegraphics[width=0.8\textwidth]{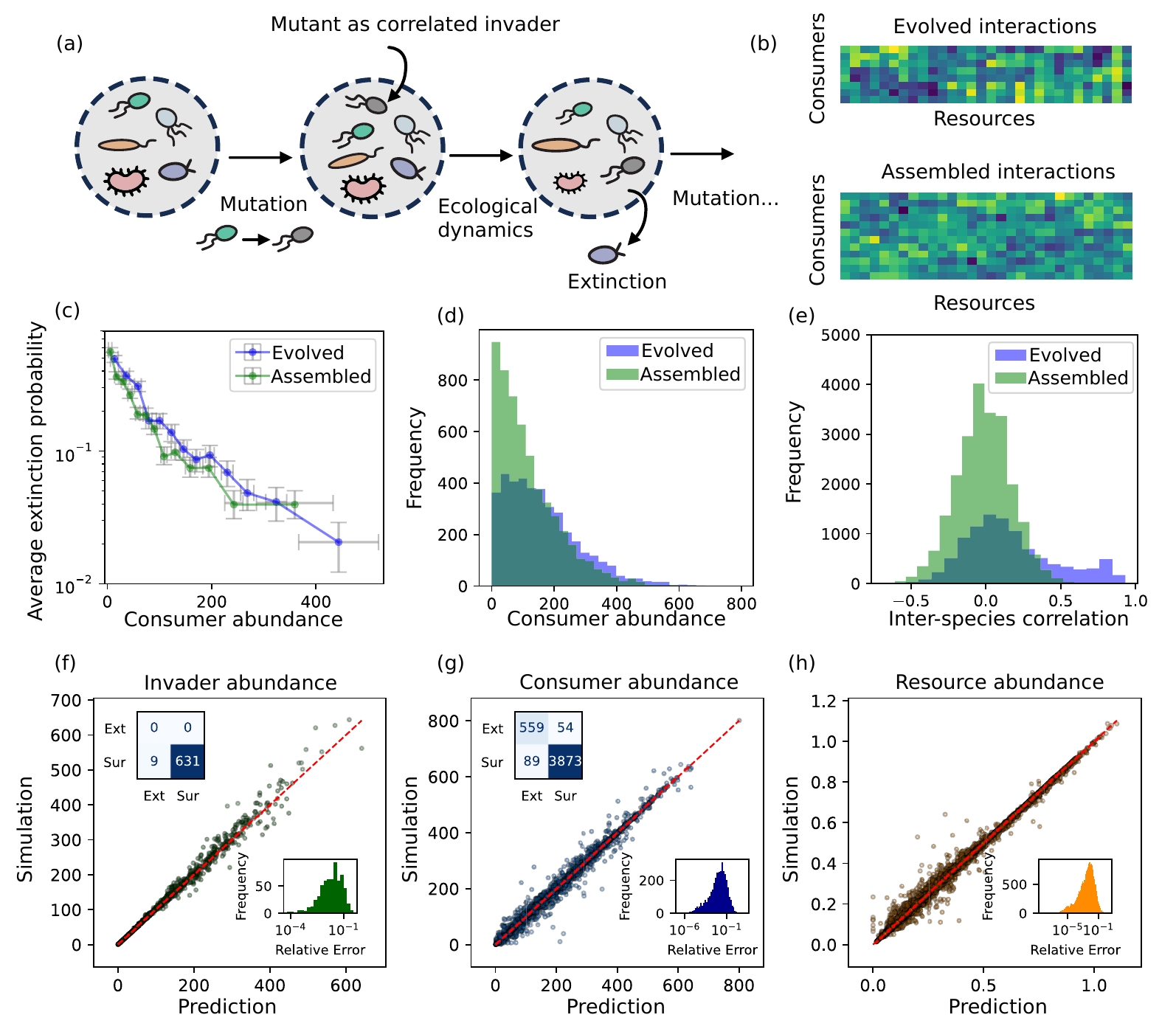}
 \caption{\textbf{Evolved communities are markedly different from assembled communities, yet predictable.} (a) Schematic showing one step of community evolution; starting with a steady-state community, we invade communities with mutants --- highly correlated with a randomly chosen parent in consumer preferences --- one at a time till communities reach a new steady state. (b) Consumer preference matrices of the evolved community after 40 evolutionary steps (top) and the original assembled community (bottom); the evolved community matrix is more structured with correlated species coexisting. (c) Extinction probability as a function of pre-invasion abundance showing an exponential decay. The error bar shows the standard error. (d) Histogram of species abundances and (e) inter-species correlations for evolved and assembled communities showing statistical differences between them. Evolution replaces uncorrelated low-abundance species with more correlated ones. (f-h) Comparison between simulated and predicted abundances for the mutant, community members, and resources in evolved communities, similar to Fig. \ref{fig:all models agree}, showing our theory still makes accurate predictions despite evolved communities having complex correlations.}
\label{fig: evolved}
\end{figure*}

A particularly interesting application of our theoretical framework is using it to understand eco-evolutionary dynamics in microbial ecosystems \cite{good2023eco}. In order to do so, we view the emergence of a mutant as an invasion by a new species that is highly correlated with an existing species in the community (the parent).  We focus on the ``strong-selection-weak-mutation''  regime \cite{gillespie1983some} in which a mutant either fixes or goes extinct before a subsequent mutation can occur. In this limit, evolution can be modeled as a sequence of invasions by successive mutations. Following each invasion, we run ecological dynamics to steady state and permanently remove any species that go extinct (see Fig. \ref{fig: evolved}(a)). For simplicity, we neglect stochastic dynamics of the mutant; this is equivalent to assuming that the community has a large population size.

Our theoretical framework is especially well-suited for analyzing short-term evolutionary dynamics of this type because it makes no assumptions about invader characteristics or the ecosystem being invaded.  This flexibility allows us to use our theoretical results to model eco-evolutionary processes despite the fact that evolution often generates complex correlations between species that co-exist in the ecosystem. An example of this is shown in Fig. \ref{fig: evolved}(b) which shows how the consumer preferences of species in the MiCRM change as a community is evolved by successive attempted invasions by highly-correlated mutants (Methods). 
Remarkably, after just $40$ successful mutations, the consumer preference matrices of the evolved and assembled communities looked qualitatively distinct (Fig, \ref{fig: evolved}(b)). 
Quantifying these differences \Akshit{revealed two key distinctions. First, species in evolved communities exhibit much higher trait correlations, with the inter-species correlation histogram showing a second peak near 0.8 (Fig. \ref{fig: evolved}(e)). This is reflected in a generic block-like structure in evolved consumer-resource interactions (Fig.~\ref{fig: evolved}(b))\cite{gralka2023genome}. Second,} evolved communities contain fewer low-abundance species (Fig. \ref{fig: evolved}(d)). This pattern emerges because low-abundance species disproportionately go extinct following successful mutations (Fig. \ref{fig: evolved}(c)), regardless of whether communities are evolved or assembled.

\begin{figure}[t]
\centering
\includegraphics[width=0.45\textwidth]{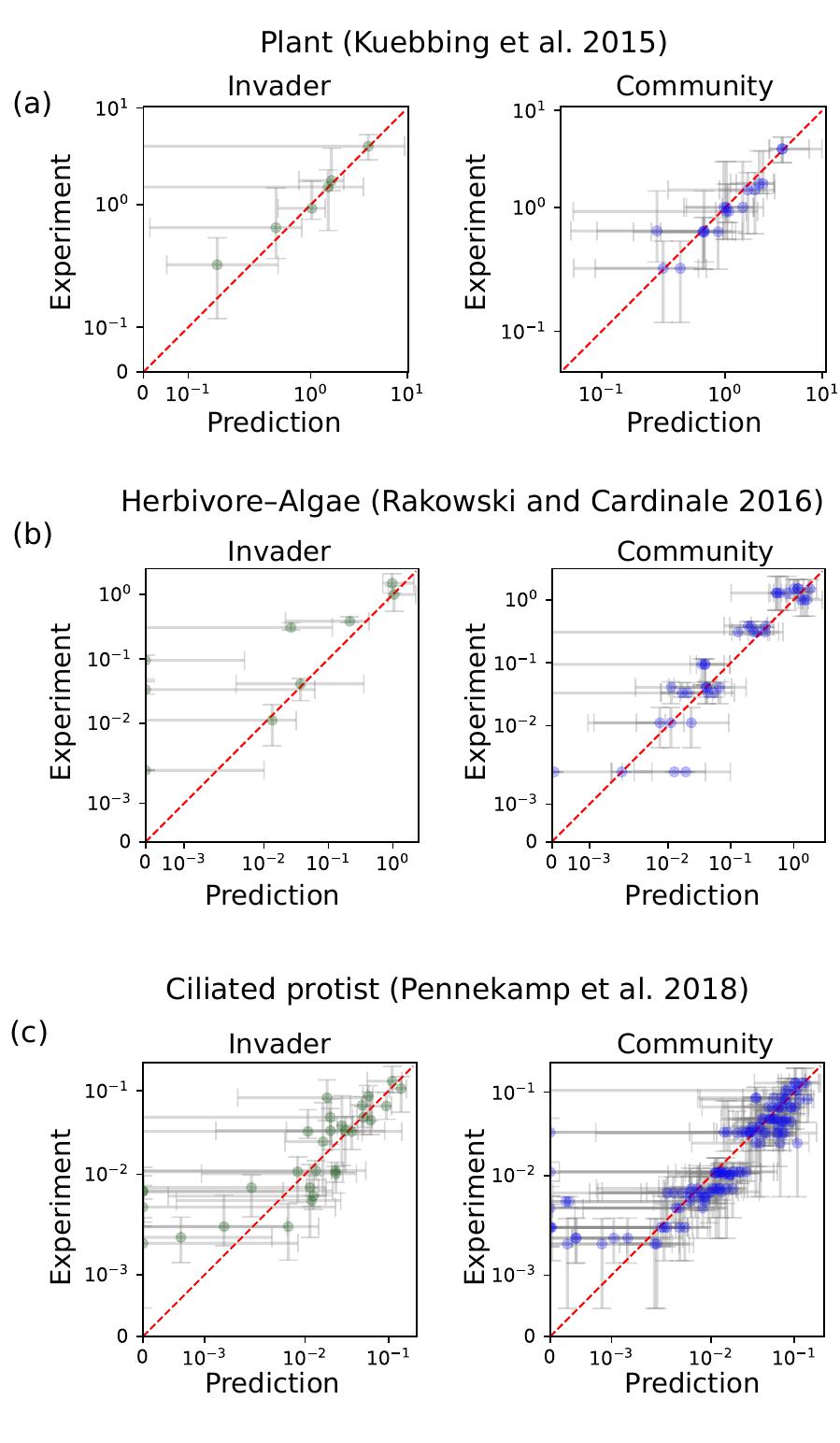}
 \caption{\textbf{Theory accurately predicts invasion outcomes across diverse experimental ecosystems.} Scatter plots comparing  experimentally measured abundances with theoretical predictions for three datasets assembling diverse ecological communities. We considered pairs of communities which differed from each other by the introduction of exactly one species, which we call the invader (Methods). Using our theory for the GLV model on these datasets, we predict both the invader abundance (left) and post-invasion community abundances (right). Points indicate median abundances for measurements, with error bars indicating variability across replicates. Both axes use a sym-log scale, with linear scaling applied below the smallest non-zero data point. The datasets span (a) plant communities \cite{kuebbing2015above}, (b) herbivore–algae communities \cite{rakowski2016herbivores}, and (c) ciliated protist communities \cite{pennekamp2018biodiversity}. 
 }
\label{fig:data}
\end{figure}

Given the very different structure of evolved and assembled communities and the fact that mutants are highly correlated with species that already exist in the community, we wanted to directly test whether our theoretical framework could be used to predict the outcome of a single step of evolution. To do so, we simulated the evolution of 640 distinct communities and then introduced a new mutant into each of these evolved communities (see Appendix \ref{app: evo numerical}). We then used our theory to predict whether the the mutant would fix, and if it fixed, the abundance of the mutant as well as the abundances of the species and resources in the community. Despite the complexities introduced by evolutionary history, Fig. \ref{fig: evolved}(f-h) shows that our theoretical predictions show remarkable agreement with simulations. These results demonstrate that evolutionary dynamics can be captured --- despite complex evolutionary histories, mutant-parent correlations, and community diversity --- as long as one incorporates ecological feedbacks.

\section{Validation using experimental data}

To test whether our framework could successfully predict invasion outcomes in \emph{stable} experimental communities, we analyzed \Akshit{six} datasets spanning diverse ecological systems (three shown in Fig.~\ref{fig:data}; remaining in Fig.~\ref{fig:extra data}). Each dataset included species abundance measurements collected under consistent environmental conditions, with different combinations of species seeded in the ecosystem. Each ecosystem contained \Akshit{4--5} species spanning different length scales and trophic levels: from plants to herbivore-algae to ciliated protists.

Even though these were not originally designed as invasion experiments, the large number of species combinations allowed us to probe invasion outcomes using a comparative approach. We took pairs of species combinations that differed from each other by the introduction of exactly one species. We then applied our framework viewing this one species as an invader. Thus, the community where this species was not introduced  was the pre-invasion community, and the one where it was introduced was the post-invasion community. The large number of species combinations and relatively low community diversity allows one to infer species interactions matrices for each ecosystem, \Akshit{done separately by fitting a GLV model to the data \cite{maynard2020predicting,lemos2024phylogeny}. In some datasets, the presence of experimental replicates allow us to estimate error bars. These experiments provide the information needed to implement our framework, and comprise an appropriate test of our theory.}

Applying our theoretical framework to all datasets, we predicted: (a) whether the invader would successfully invade; if yes,  the post-invasion (b) invader abundance and (c) resulting shift in community abundances (for details, see Methods in Appendix I). Fig. \ref{fig:data} shows results for three datasets: (a) plant communities from the southeastern United States (data from Kuebbing et al. \cite{kuebbing2015above}); (b) herbivore-algae communities (data from Rakowski and Cardinale \cite{rakowski2016herbivores}); and (c) ciliated protist communities grown across a range of temperatures (15°C to 25°C; data from Pennekamp et al \cite{pennekamp2018biodiversity}). \Akshit{Fig.~\ref{fig:extra data} shows three additional plant datasets from a series of distinct biodiversity-ecosystem functioning (BEF) experiments \cite{tilman2001diversity,van2010diversity, cadotte2013experimental}.} We found that predictions broadly agreed with measurements across all datasets. Specifically,  while the median of our predictions agreed with the median of measurements, the size of the error bars varied across datasets. Generally, prediction errors decreased with the number of replicates, \Akshit{suggesting that experimental noise in the datasets was the dominant source of error}. Errors were smallest for the plant community dataset (Fig. \ref{fig:data}a), which had the largest number of replicates (10); and \Akshit{highest} for herbivore-algae communities (Fig. \ref{fig:data}b) with the lowest number of replicates (5). \Akshit{The additional datasets (Fig.~\ref{fig:extra data}) had no replicates, so we could not estimate error bars for them. } Further, deviations from measurements were more likely for species with low abundances (e.g., see points on the bottom-right of both panels of Fig. \ref{fig:data}c). \Akshit{These deviations stem from several sources of error: primarily measurement errors and biological variability, alongside uncertainty in fitted interaction strengths, as well as GLV dynamics being an approximation of the true dynamics. By inferring invasion fitness using GLV models fitted for these datasets (Methods), we also confirmed that dressed invasion fitness is a much better predictor of invader abundance than naive invasion fitness (Fig.~\ref{fig: data dressed invasion fitness} and Fig.~\ref{fig: extra data dressed invasion fitness}). Taken together, these results demonstrate that our framework can not only predict, but also provide insight into invasion outcomes for diverse ecosystems spanning a range of environments, size scales, and trophic levels.}

\section{Discussion}
In this paper, we develop a theory for predicting the outcomes of ecological invasions applicable to a wide variety of ecological models. Our framework treats invasions in diverse ecosystems as perturbations to existing communities, allowing us to predict which species go extinct, how surviving species' abundances shift, and the invader's final abundance. We validate our predictions through both numerical simulations and analyses of experimental datasets spanning plant, herbivore-algae, and protist communities. Our theoretical framework is predictive across models with fundamentally different mathematical structures---from Lotka-Volterra to consumer resource models with and without cross-feeding---and across systems of different sizes---from small experimental communities with 4--5 species to large simulated ecosystems with 10s to 100s of species. This broad applicability lies in the generality of viewing invaders as perturbations near a steady state and the power of a linear response theory augmented with discontinuous extinction events.

A fundamental insight from our theory is the identification of a ``dressed invasion fitness'' as the central quantity governing invasion outcomes. Unlike conventional ``naive'' invasion fitness \cite{chesson1990macarthur,te2013invasion} which only measures the initial growth rate of an invader in the resident community, the dressed invasion fitness incorporates ecological feedbacks between the invader and the resident community. We show that this distinction is crucial for accurately predicting both invasion success and the resulting community structure. This is because mathematically, the effect of an invader abundance decomposes into a contribution proportional to its dressed invasion fitness and a contribution from invasion-induced extinctions. Both these contributions incorporate ecological feedbacks between the invader and the full community into which it invades. \Akshit{Similar feedbacks appear in the cavity method for models of ecosystems, but these approaches are completely different in scope. Our framework uses known interactions to make predictions for specific invasions in specific communities, while the cavity method uses random parameters and can only describe statistical properties of an ensemble of invasions.}

A direct implication of our framework is that eco-evolutionary dynamics in complex communities follow a pattern where successful mutants predominantly replace low-abundance community members rather than their parent species \cite{mcenany2024predicting}. This stands in contrast to the canonical view from population genetics --- where displacement occurs among closely related lineages --- as well as pairwise niche overlap metrics \cite{shea2002community}. \Akshit{While parents and mutants do compete more strongly among themselves than with other community members, the cumulative effect of weaker interactions across a sufficiently diverse community may ultimately also affect the fate of a mutant. In our work, this occurs because mutants and parents interact not only directly with each other, but also through global feedbacks mediated by the entire community structure. This contrasting perspective motivates further empirical tests as well as theoretical work on population genetics in complex ecological communities. After successful invasion, the extinction probabilities of other residents decay exponentially with their pre-invasion abundance}, effectively shielding abundant species while rendering rare species vulnerable regardless of their relatedness to the invader. \Akshit{These patterns are consistent with observations from natural communities which show a decreasing pattern of coexistence with phenotypic similarity \cite{lemos2024phylogeny,sireci2023environmental}. Our results augment these observations: despite the decrease, in a diverse community, phenotypically similar species are still reasonably likely ($\sim 40\%$) to coexist (Fig.~\ref{fig:extinction versus correlation}).} This observation \Akshit{also} highlights how ecological feedbacks fundamentally reshape evolutionary trajectories in ways that cannot be captured by traditional models that ignore invader-induced feedback.

Our theory also helps explain qualitative differences in invasion outcomes across different ecological models. In the Lotka-Volterra model with competitive interactions, species extinctions consistently increase overall community abundance because extinctions reduce negative interactions caused by competition. In contrast, extinctions in consumer-resource models can cause the abundance of surviving species to both increase or decrease. This difference arises because in consumer-resource models, interactions among consumers are mediated by resources, allowing for more diverse outcomes through complex resource-mediated feedback mechanisms.

In the future, it would be interesting to extend our theoretical framework to understand invasions in communities that exhibit time-varying dynamics such as limit cycles or chaos. Recent work shows that even in chaotic regimes, abundant species reach long-lived transient states that share many properties with steady states \cite{arnoulx2024many,blumenthal2024phase}. This suggests that our framework might be adaptable to predict outcomes in dynamic regimes by viewing chaos as a series of epochs where the community is temporarily close to some steady state. Another promising direction would be to apply and extend our framework to community coalescence in which entire communities merge \cite{rillig2015interchange,diaz2022top,custer2024toward}. Our current theory assumes single-species invasions, whereas coalescence involves many simultaneous invaders.

Finally, while our work focuses primarily on short-term eco-evolutionary dynamics in the strong-selection-weak-mutation regime, it would be interesting to extend the framework to incorporate statistical predictions for long-term evolutionary dynamics. This would provide valuable insights into how ecological feedbacks shape community structure over long evolutionary time. The mathematical tools developed here provide a foundation for addressing these more complex scenarios while maintaining the conceptual clarity offered by linear response and perturbation theory.

\section*{Methods}
Please see SI Appendices for detailed Materials and Methods. These appendices contain: (1) derivations for our theoretical framework, including invasion outcomes for a variety of ecological models and all equations in the main text; (2) algorithms for implementing our theory for specific models, used as tests in Figs. 2--3; (3) methods for all simulations performed and parameters used; (4) extensions of our framework and generalization to multiple invaders and sources of environmental perturbations.

\section*{Acknowledgments}
We thank Shing Yan Li for discussions. E.B. acknowledges support from the National Science Foundation Graduate Research Fellowship Program and the Fannie and John Hertz Foundation. This work was funded by NIH NIGMS R35GM119461 to P.M. and Chan-Zuckerburg Institute Investigator grant to P.M. A.G. acknowledges support from the Ashok and Gita Vaish Junior Researcher Award, the DST-SERB Ramanujan Fellowship, as well the DAE, Govt. of India, under project no. RTI4001.

\bibliography{perturbation}

\clearpage
 \newpage

\onecolumngrid
\appendix

\begin{center}
\Large\textbf{Appendices for \\A theory of ecological invasions and its implications for eco-evolutionary dynamics}
\end{center}

\setcounter{figure}{0}
\renewcommand{\thefigure}{S\arabic{figure}}

\section{Derivation of linear response theory for the Generalized Lotka-Volterra (GLV) models}
In this section, we provide a detailed derivation of Eq. \eqref{eq:community shift} which relates community shift to invader abundance and the pre-invasion abundance of extinct species. Then we derive Eq. \eqref{eq:invader abundance}, which predicts invader abundance given the knowledge of pre-invasion community, invader-community interaction, and the knowledge of what survives after invasion. Finally, in the last subsection we go beyond the scope of the main text and extend the framework to allow multiple invaders and environmental perturbations. 

\subsection{Primer: Environmental perturbations without extinctions}
A simple primer to consider before considering invasion and extinction is when environmental perturbations alter species abundance without causing extinctions. The steady state before invasion $X_j^{\text{old}}$ is given by linear equality $r_i-\sum_{j}A_{ij}X_j^{\text{old}}=0$, solving the linear equation leads to 
\begin{align}
    X_i^{\text{old}}&=A_{ij}^{-1}r_j.
\end{align}
Now suppose there is an environmental perturbation $r_i\to r_i+\delta r_i$, the new steady state condition is simply $r_i+\delta r_i-\sum_{j}A_{ij}X_j^{\text{old}}=0$, leading to new abundance
\begin{align}
    X_i^{\text{new}}&=A_{ij}^{-1}(r_j+\delta r_j).
\end{align}
Therefore, the change in steady state abundance is simply
\begin{align}
    \delta X_i^{\text{no extinction}}=A_{ij}^{-1}\delta r_j.
    \label{eq: appendix env perturb}
\end{align}
Here this relationship is exact instead of perturbative; the steady-state condition for surviving species in the Lotka-Volterra model is exactly linear.

\subsection{Derivation of Eq. \eqref{eq:community shift}: Invasions and extinctions as effective perturbations}
In this work, instead of actual environmental perturbation, we make use of the insight that in a background of large community, a species feels the addition of a new interaction or a deletion of an old interaction no differently than an environmental perturbation. More explicitly, solving Eq. \eqref{eq: surviving equality} leads to 
\begin{align}
    \vec{r_\Surviving}-A_{\Surviving\Surviving}\vec{X}^{\text{new}}_\Surviving-A_{\Surviving0}{X}_0&=0,\\
\vec{X}^{\text{new}}_\Surviving&=A_{\Surviving\Surviving}^{-1}(\vec{r_\Surviving}-A_{\Surviving0}{X}_0).
\end{align}

At the same time, the old steady state condition for these surviving species is 
\begin{align}
    \vec{r_\Surviving}-A_{\Surviving\Surviving}\vec{X}^{\text{old}}_\Surviving-A_{\Surviving\Extinct}\vec{X}_\Extinct&=0,\\
\vec{X}^{\text{old}}_\Surviving&=A_{\Surviving\Surviving}^{-1}(\vec{r_\Surviving}-A_{\Surviving\Extinct}\vec{X}_\Extinct).
\end{align}
Therefore, their change in abundance is
\begin{align}
\delta \vec{X}_\Surviving&=\vec{X}^{\text{new}}_\Surviving-\vec{X}^{\text{old}}_\Surviving,\\
&=A_{\Surviving\Surviving}^{-1}(-A_{\Surviving0}{X}_0+A_{\Surviving\Extinct}\vec{X}_\Extinct).
\end{align}
This recovers Eq. \eqref{eq:community shift}. Moreover, comparison with Eq. \eqref{eq: appendix env perturb} shows the role of effective environmental perturbation played by invader and extinct species influence, $\vec{r}^\text{eff}=-A_{\Surviving0}{X}_0+A_{\Surviving\Extinct}\vec{X}_\Extinct$.

\begin{figure*}[!ht]
\centering
\includegraphics[width=1\textwidth]{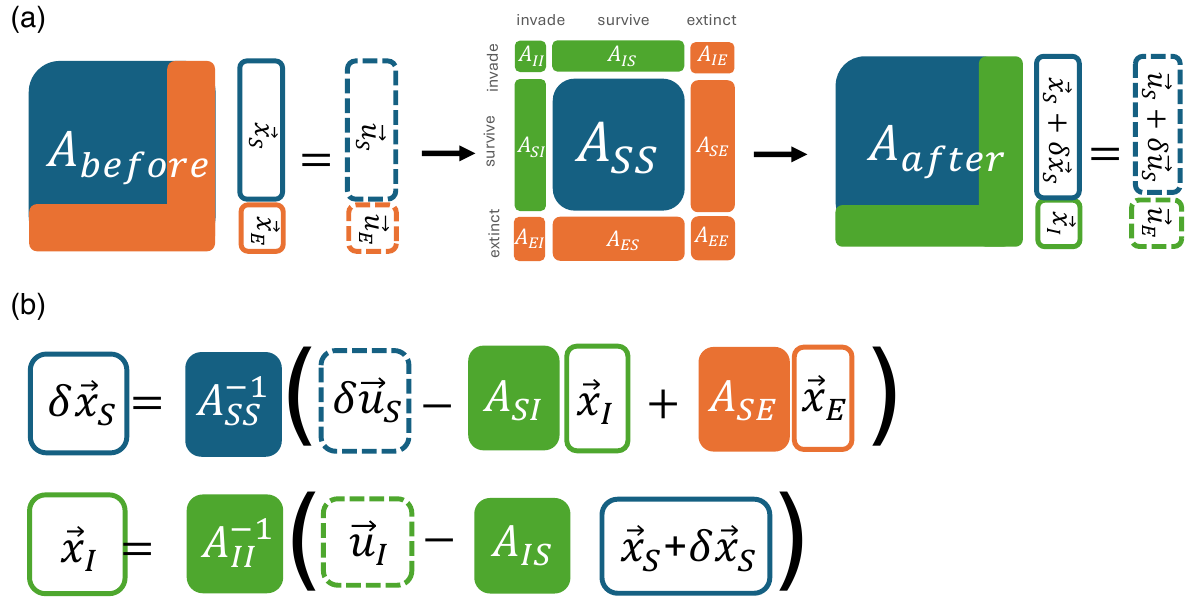}
 \caption{ \textbf{Equations showing how our theory  partitions an ecosystem into invaders, survivors and extinct species.} (a) shows how the structure of interactions changes in the Lotka-Volterra model before and after an invasion, and (b) shows how the community shifts and invader abundance is related to each part of the system shown in Eq. \eqref{eq: 9 blocks}  }
\label{fig:euqation figures}
\end{figure*}

\subsection{Derivation of Eq. \eqref{eq:invader abundance}: Predicting invader abundance} 
\label{app: derive invader abundance}
The invader introduces an extra constraint on the steady state condition, namely 
\begin{align}
    r_0-A_{0\Surviving}\vec{X}_\Surviving^{\text{new}}-A_{00}X_0&=0\\
     r_0-A_{0\Surviving}\vec{X}_\Surviving^{\text{old}}-A_{0\Surviving}\delta \vec{X}_\Surviving-A_{00}X_0&=0
\end{align}
We can then substitute Eq. \eqref{eq:community shift} to obtain

\begin{align}
     r_0-A_{0\Surviving}\vec{X}_\Surviving^{\text{old}}-A_{0\Surviving}A_{\Surviving\Surviving}^{-1}(-A_{\Surviving0}{X}_0+A_{\Surviving\Extinct}\vec{X}_\Extinct)-A_{00}X_0&=0
\end{align}
Solving for $X_0$ arrives at 
\begin{align}
     X_0=\frac{r_0-A_{0\Surviving}\vec{X}_\Surviving^{\text{old}}-A_{0\Surviving}A_{\Surviving\Surviving}^{-1}A_{\Surviving\Extinct}\vec{X}_\Extinct}{A_{00}-A_{0\Surviving}A^{-1}_{\Surviving\Surviving}A_{\Surviving0}},
\end{align}
which recovers Eq. \eqref{eq:invader abundance}

\subsection{Extended framework for multiple invaders and environmental perturbations}
To extend our theory to accommodate multiple invaders, we need to separate indices of all species into those invading ($\Invading$), extinct ($\Extinct$) and surviving ($\Surviving$), illustrated in Fig. \ref{fig:euqation figures}. In explicit block matrix notation, the interaction matrix of the new system consists of 9 block matrices,
\begin{align}
A=
\begin{pmatrix}
A_{\Surviving\Surviving}&A_{\Surviving\Invading}&A_{\Surviving\Extinct}\\
A_{\Invading\Surviving}&A_{\Invading\Invading}&A_{\Invading\Extinct}\\
A_{\Extinct\Surviving}&A_{\Extinct\Invading}&A_{\Extinct\Extinct}
\end{pmatrix}.
\label{eq: 9 blocks}
\end{align}

This block structure and relationship with the self-consistency equations is illustrated in the diagram in Fig. \ref{fig:euqation figures} (b).

In general, any perturbation to the system is composed of 3 parts, including direct environmental perturbation, invasions, and extinctions respectively. These effects can all be absorbed into an effective environmental perturbation $\vec{\delta r}^{(eff)}$,
\begin{align}
    \delta \vec{r}^{\text{eff}}= \delta \vec{r}+\delta \vec{r}^{\text{inv}}+\delta \vec{r}^{\text{ext}},
\end{align}
where 
\begin{equation}\vec{r}^{\text{inv}}=-A_{\Surviving\Invading} \vec{X}_{\Invading}, \quad \vec{r}^{\text{ext}}=A_{\Surviving\Extinct} \vec{X}_{\Extinct}\end{equation}

The corresponding linear perturbations in abundance, the generalized version of Eq. \eqref{eq:community shift}, is 
\begin{align}
    \delta \vec{X}_{\Surviving}=A^{-1}_{\Surviving\Surviving}\delta \vec{r}^{(eff)}_{\Surviving}= \delta \vec{X}_{\Surviving}&=A^{-1}_{\Surviving\Surviving}( \delta \vec{r}_{\Surviving} -A_{\Surviving\Invading} {\vec{X}}_{\Invading}+A_{\Surviving\Extinct} {\vec{X}}_{\Extinct}).
     \label{eq:deltax}
\end{align}

At the same time, abundance of surviving invaders needs to follow the set of new steady state conditions
\begin{align}
    r_0-A_{0\Surviving}\vec{X}_\Surviving^{\text{new}}-A_{\Invading\Invading}\vec{X}_\Invading&=0
    \label{eq:invaderx}
\end{align}

Substituting \eqref{eq:deltax} into \eqref{eq:invaderx} like before, we obtain 
the generalized version of Eq. \eqref{eq:invader abundance}

\begin{align}
    \vec{X}_{\Invading}  &= \mathbf{M}_{\Invading\Invading}^{-1} \left( \vec{f}_I+ A_{\Invading\Extinct} \vec{X}_{\Extinct} - A_{\Invading\Surviving} A_{\Surviving\Surviving}^{-1} \delta \vec{r}_{\Surviving} - A_{\Invading\Surviving} A_{\Surviving\Surviving}^{-1} A_{\Surviving\Extinct} \vec{X}_{\Extinct} \right),
    \label{eq: general XI}
\end{align}
where \begin{align}
\mathbf{M}_{\Invading\Invading}& = A_{\Invading\Invading} - A_{\Invading\Surviving} A_{\Surviving\Surviving}^{-1} A_{\Surviving\Invading},\\
\vec{f_I}&=\vec{r}_{\Invading} - A_{\Invading\Surviving} \vec{X}_{\Surviving}-A_{\Invading\Extinct} \vec{X}_{\Extinct}.
\end{align}

\section{Predicting invasion in the presence of multi-stability}
Our framework can predict the correct invasion outcome even if the dynamical system has multiple stable fixed points, given that identity of surviving species are not significantly changed. We demonstrate this using Lotka-Volterra model with high enough variance in interaction strength. It has been shown that a phase transition change the system from unique stable steady states to multiple stable steady states \cite{bunin2017ecological} in random Lotka-Volterra models when the variance of insteraction increases. We demonstrate this by explicitly sampling many initial state to check if the system exhibit multiple fixed points, then estimates the portion of steady-states exhibiting multistability given the level of variance. Finally we choose a high enough variance that guarantee multi-stability happens more than 80 percent of the time. 

\begin{figure*}[!h]
    \centering
    \includegraphics[width=0.7\textwidth]{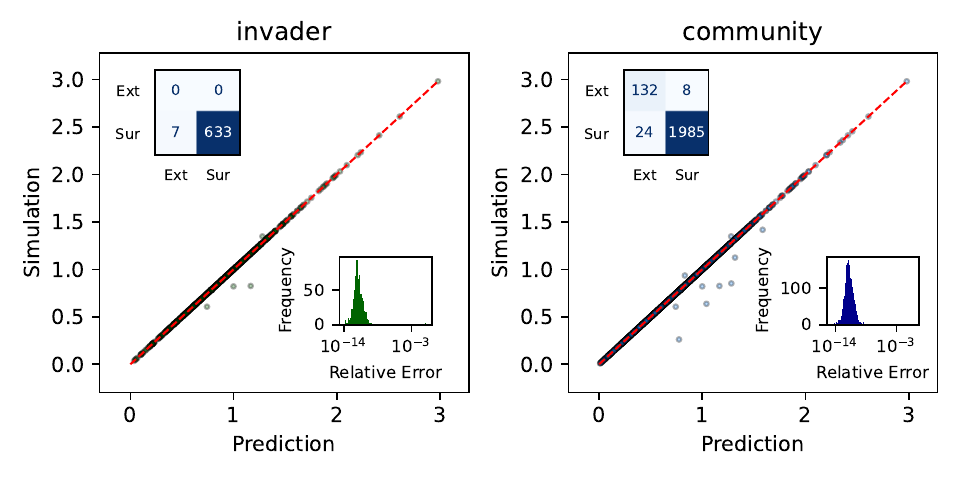}
    \caption{ \textbf{Comparison of simulation outcomes and theoretical predictions for Lotka-Volterra models exhibiting multiple stable equilibria prior to invasion.} The resident community's interaction matrix is sampled with high variance ($\sigma_c = 3.16$), generating systems with multiple locally stable fixed points identified by simulating dynamics from diverse initial conditions. Invader--community interaction strengths are drawn from a distribution with lower variance ($\sigma_{\text{inv}} = 0.316$), ensuring that invaders exert a limited perturbative effect on the resident community.}
    \label{fig:multistability_combined}
\end{figure*}

In the main text, we sample the invader using the same distribution as that of the regional species pool, which forms the resident community before invasion. This choice models the natural occurrence of invaders originating from a statistically similar pool of species. However, when we increase the variance of interactions in the regional species pool, some interactions become negative, representing ``cooperative'' rather than competitive interactions. This cooperation allows species from the pool to become large-impact invaders that can potentially drive most resident species to extinction. While the self-consistency equation remains exact in such situations, the iterative scheme assumes that a considerable fraction of species survive after invasion. Therefore, in the demonstration in Fig.~\ref{fig:multistability_combined}, we reduced the variance of the interaction coefficients for the pool from which we sample invaders. In this simulation, we set the mean interaction to be \(\mu_A = 0.1\), the variance of the regional species pool for the resident community as \(\sigma_A = 3.16\), and the variance for sampling invaders as \(\sigma_A^{\text{invader}} = 0.316\). The size of the regional species pool is 10, with approximately 3 species surviving in the resident community. This demonstrates that the self-consistency equations and the iterative approach remain applicable in the presence of multiple steady states.

\section{General linear response theory}
\label{app:general framework}
We propose the following generalized form of ecological models:
\begin{align}
    \frac{dX_i}{dt} = f_i(X_i)g_i(\vec{X}, \vec{r}),
    \label{eq: general ode}
\end{align}
where $X_i$ represents the abundance of each species, and $\vec{r}$ parameterizes the environmental influences on growth. The function $f_i(X_i)$ governs species extinction dynamics. It is positive for surviving species, and it captures the concept of extinction for biotic species as $f_i(X_i)$ must satisfy 
    \begin{align}
        f_i(0) = 0.
    \end{align}
It simply means that biotic species cannot grow again once their abundance reaches zero, as biotic species grow by reproduction, an organic self-replication. In most models, $f_i(X_i)$ is simply $X_i$. Instead, for abiotic species like nutrients or metabolites that grow through supply, extinction should not apply, and we can simply set $ f_i(x) = 1$.

In contrast, the function $g_i(\vec{X}, \vec{r})$ governs species growth and interactions. The steady-state condition for species abundance $\vec{X}^*$ is determined by:
\begin{align}
    g_i(\vec{X}^*, \vec{r}) \begin{cases} =0 &\text{if species $i$ survives}   \\
    < 0 &\text{if species $i$ becomes extinct}
    \end{cases},
    \label{eq:old steady state}
\end{align}
where the second condition corresponds to the stability to re-introduction by extinct species. This tells us that instead of linearizing the complete dynamics in Eq.\eqref{eq: general ode}, it is more natural to linearize the steady-state condition if we consider perturbations to the ecosystem. Note that the full dynamics are \textit{always nonlinear} because of the presence of $f_i(X_i)$. When we refer to linearization in this appendix, we refer to approximating $g_i$ as a linear function by expanding it around a given steady-state set of abundances $\vec{X}$ and growth rates $\vec{r}$. As we show below, this approximation is mathematically equivalent to assuming that the effective interaction matrix strengths do not change too much between the pre- and post-invasion communities. We might refer to this general class of models with nonlinear $g_i$ as  ``nonlinear ecological models'' for brevity; to clarify all models considered in this manuscript are nonlinear in their full dynamics. We merely mean that the per-capita growth rates are nonlinear when we refer to this shorthand.

We first study the scenario where all species survive, following the first condition in Eq.\eqref{eq:old steady state}. We want to study the abundance change to each species,
$\vec{X} \to \vec{X}+\delta \vec{X}$, in response to perturbation of environmental parameter $\vec{r}\to \vec{r}+ \delta \vec{r}$. The new steady state needs to satisfy 
\begin{align}
    g_i(\vec{X}+\delta \vec{X},r_i+\delta r_i)=0.
    \label{eq:new steady state}
\end{align}
Define the effective interaction matrices as the derivatives of the non-zero steady state conditions,  
\begin{align}
    \mathcal{A}_{ij} &= \left.\frac{\partial g_i(\vec{X}, \vec{r})}{\partial X_j}\right|_{\vec{X} = \vec{X}^\text{old}}, \quad 
    \mathcal{B}_{ij} = \left.\frac{\partial g_i(\vec{X}, \vec{r})}{\partial r_j}\right|_{\vec{X} = \vec{X}^\text{old}}
    \label{eq:SI effective A definition}
\end{align}
Note that in general effective interactions \(\mathcal{A}\) and \(\mathcal{B}\) depend on \(\vec{X}\), but assuming they remain constant at the new steady state \(\vec{X}^*\) is precisely the same as performing a first-order Taylor expansion of the new steady state condition in Eq.~\eqref{eq:new steady state}. This linear approximation of the change in the steady state function around \(\vec{X}^*\) yields
\begin{align}
    \mathcal{A} \delta \vec{X} + \mathcal{B} \delta \vec{r} = 0.
\end{align}
By simplifying and labeling 
\[
    A^{-1} = -\mathcal{A}^{-1} \mathcal{B},
\]
the abundance shift satisfies the linear approximation
\begin{align}
    \delta \vec{X} = A^{-1} \delta \vec{r}.
    \label{eq: simplified general perturbation}
\end{align}
Here, the linearity expressed through the derivatives \(\mathcal{A}\) and \(\mathcal{B}\) corresponds exactly to the statement that changes in \(\vec{X}\) can be well approximated by applying these effective interaction matrices to changes in \(\vec{r}\), illustrating the equivalence between locally linear functions and their differential approximations.

In the context of invasion theory, we have effective environmental perturbation $ \delta \vec{r}^{\text{eff}}$ consists of invasion, extinction and actual environmental perturbation, the same way as linear model in Eq.\eqref{eq: 9 blocks} and Fig. \ref{fig:euqation figures}:
\begin{align}
    \delta \vec{r}^{\text{eff}}= \delta \vec{r}+\delta \vec{r}^{\text{inv}}+\delta \vec{r}^{\text{ext}},
\end{align}
where
\begin{equation}\delta \vec{r}^{\text{inv}}=-{A}_{\Surviving\Invading} \vec{X}_{\Invading}, \quad \delta \vec{r}^{\text{ext}}={A}_{\Surviving\Extinct} \vec{X}_{\Extinct}.\end{equation}
Therefore the abundance shift to surviving species is 
\begin{align}
    \delta X={A}^{-1}_{\Surviving\Surviving}(-{A}_{\Surviving\Invading} \vec{X}_{\Invading}+{A}_{\Surviving\Extinct} \vec{X}_{\Extinct}+\delta \vec{r})
\end{align}

To summarize, to predict the linear response of the steady state in any nonlinear model, the first step is to compute the effective interaction matrix, and compute the matrix inverse given by $A^{-1}$. Once this matrix is determined, the problem simplifies and becomes equivalent to the generalized Lotka-Volterra model. This simple exercise of Taylor expansion justifies the use of the Lotka-Volterra model despite its simplicity - it captures the linear response of ecosystems near their steady states. Finally, the result maps to the one derived for the generalized Lotka-Volterra model in Eq.\eqref{eq:community shift}. 

A similar derivation for Eq.\eqref{eq:invader abundance} will require solving only the equation for invader steady state abundance, i.e., $g_{i\in \Invading}(\vec{X}^{\text{new}},r_i)=0$. 
\begin{align}
  g_{i\in \Invading}(\vec{X}_\Surviving^{\text{old}}+\delta\vec{X},\vec{X}_\Extinct,X_i,r_i)&=0,  \\
  \sum_{j\in \Surviving} \frac{\partial g_{i\in \Invading,j}}{\partial  X_j}\delta {x_j}+\sum_{k\in \Invading} \frac{\partial g_{i\in \Invading,k}}{\partial  X_k}{x_k}+g_{i\in \Invading}(\vec{X}^{\text{old}}_\Surviving,\vec{X}^{\text{old}}_\Extinct=0 ,X_i=0,r_i)&=0.
\end{align}
The last part of the expression corresponds almost exactly to the naive invasion fitness, except that contributions from extinct species are set to zero. This portion uniquely captures the nonlinear effects present in the growth rate, whereas the first two terms arise purely from the linear response of the system.

In matrix notation, it becomes simply
\begin{align}
    {A}_{IS}\delta {\vec{X}_S}+{A}_{\Invading\Invading}\vec{X}_\Invading+\vec{g}_{ \Invading}(\vec{X}^{\text{old}},X_i=0,r_0)&=0.
\end{align}

Therefore, the invader abundance formula in terms of dressed invasion fitness Eq.\eqref{eq:x0interp} applies even for nonlinear consumer growth rate.

\section{Extinct species have negative invasion fitness}
If we do a perturbation around a stable steady state by adding a little bit of the extinct species $i$, the new growth rate follows 
\begin{equation}\frac{d}{dt}(x_i+\delta x_i)=(x_i+\delta x_i)(k_i-\sum_{j\neq i}A_{ij}x_j-A_{ii}\delta x_i)\end{equation}
for extinct species, $x_i=0$, thus 
\begin{equation}
    \frac{d\delta x_i}{dt}=\delta x_i(k_i-\sum_{j\neq i}A_{ij}x_j-A_{ii}\delta x_i)
\end{equation}
The terms in the bracket are almost the invasion fitness. It needs to be negative for a stable fixed point, i.e.,
\begin{equation}(k_i-\sum_{j\neq i}A_{ij}x_j)<A_{ii}\delta x_i.
\end{equation}
For a general model, the perturbation to extinct species is
\begin{equation}\frac{d\delta X_i}{dt}=g_i(\vec{X}, \vec{r}) \frac{\partial f_i}{\partial X_i} \delta X_i 
+ f_i(X_i)  \frac{\partial g_i}{\partial X_i} \delta X_i\end{equation}
The analogous inequality is 
\begin{equation}\left. g_i(\vec{X}, \vec{r}) \frac{\partial f_i}{\partial X_i}\right|_{X_i=0} < 0  
+ \left.f_i(X_i)  \frac{\partial g_i}{\partial X_i}\right|_{X_i=0} < 0\end{equation}
But $f_i(0)=0$ by definition, and we should expect $\left.\frac{\partial f_i}{\partial X_i}\right|_{X_i=0}>0$ because it needs to allow growth for positive abundance. Therefore, for the general model, an analogous quantity to invasion fitness is simply
$\left. g_i(\vec{X}, \vec{r}) \right|_{X_i=\delta X_i} $

\section{Derivation of linear response in CRMs}
\label{app: derive CRM response}
Notice that the consumer resource model can be reformulated in a matrix form 
\begin{equation}
\begin{pmatrix}
\frac{d\vec{N}}{dt} \\
\frac{d\vec{R}}{dt}
\end{pmatrix}
=\begin{pmatrix}
\vec{N}\\
\vec{R}
\end{pmatrix}\odot\left[\begin{pmatrix}
-\vec{m}\\
\vec{K}
\end{pmatrix}-\begin{pmatrix}
0&-C \\
E^{\operatorname{T}}& Q
\end{pmatrix}\begin{pmatrix}
\vec{N}\\
\vec{R}
\end{pmatrix}\right]
\end{equation}

Comparing with the Lotka-Volterra model in matrix form,
\begin{equation}
\frac{d\vec{X}}{dt} 
=\vec{X}\odot(\vec{r}-A\vec{X})
\end{equation}
We have the mapping
\begin{align}
\vec{X}= \begin{pmatrix}
    \vec{N}\\\vec{R}
\end{pmatrix}, \quad \vec{r}= \begin{pmatrix}
    -\vec{m}\\\vec{K}
\end{pmatrix},\quad
    {A}=\begin{pmatrix}
0&-C \\
E^{\operatorname{T}}& Q
\end{pmatrix}
\end{align}

For the surviving subset of species, interaction matrices are invertible, and can be derived in block matrix form
\begin{align}
    {A}^{-1}_{\Surviving \Surviving}&=-
    \begin{pmatrix}
 -I & B^{-1}C\\
Q^{-1}E^{\operatorname{T}} & I-P
\end{pmatrix}_{\Surviving \Surviving}
\begin{pmatrix}
 B^{-1} & 0\\
0 & Q^{-1}
\end{pmatrix}_{\Surviving \Surviving}\\
    &=
    \begin{pmatrix}
      B^{-1} & B^{-1} C Q^{-1}\\
    Q^{-1}E^{\operatorname{T}} B^{-1}&  (I-P) Q^{-1} 
    \end{pmatrix}_{\Surviving \Surviving}
\end{align}

where $B$ and $P$ are defined to simplify the expressions
\begin{align}
    B=CQ^{-1}E^{T}, \quad P=Q^{-1}E^TB^{-1}C.
\end{align}

For consumer invader, we have the invader 
\begin{equation}X_0= \begin{pmatrix}
    \vec{N_0}\\0
\end{pmatrix}, \quad \vec{r_0}= \begin{pmatrix}
    -m_0\\0
\end{pmatrix},  \quad {A}_{0S/E}=\begin{pmatrix}
 0&
-C_{0S/E}
\end{pmatrix},\quad {A}_{S/E0}=\begin{pmatrix}
 0\\
E_{S/E 0}^{\operatorname{T}}
\end{pmatrix}\end{equation}

We can substitute these block-form Lotka-Volterra quantities to Eq. \eqref{eq:invader abundance} and Eq. \eqref{eq:community shift} to obtain
\begin{align}
    A_{0\Surviving}A_{\Surviving \Surviving}^{-1}A_{\Surviving 0}&=C_{0\Surviving} (1-P)Q_{\Surviving \Surviving}E_{\Surviving 0}^{\operatorname{T}},\\
    A_{0\Surviving}A^{-1}_{\Surviving \Surviving}A_{\Surviving \Extinct}&={C}_0E^{T}B^{-1}C_{\Surviving \Extinct}\vec{R}_{\Extinct}-{C}_{0S}(1-P)Q_{\Surviving \Extinct}\vec{N}_{\Extinct},\\
    A_{\Invading \Extinct } x_{\Extinct}&= -C_{0\Extinct}R_{\Extinct},
\end{align}
 which pieces together the invader abundance
\begin{equation}
    N_0=\frac{f_0-C_{0\Extinct}R_\Extinct-{C}_{0\Surviving}E_{\Surviving \Surviving}^{T}B^{-1}C_{\Surviving \Extinct}\vec{R}_{\Extinct}+{C}_{0\Surviving}P Q_{\Surviving \Extinct}\vec{N}_{\Extinct}}{C_{0\Surviving} (1-P)Q^{-1}_{\Surviving \Surviving}E_{\Surviving 0}^{\operatorname{T}}}
    \label{eq: MCRM invader abundance}
\end{equation}
where invader fitness is $f_0=C_{0\Extinct} R_{\Extinct}+C_{0\Surviving} R_{\Surviving}-m_0 $
And the abundance shifts to consumers and resources are

\begin{align}
\begin{pmatrix}
         \delta N_{\Surviving} \\
         \delta R_{\Surviving}
    \end{pmatrix}&=
    \begin{pmatrix}
      B^{-1} & B^{-1} C Q^{-1}\\
   Q^{-1} E^{\operatorname{T}} B^{-1}&  (I-P)Q^{-1} 
    \end{pmatrix}_{\Surviving\Surviving}
    \left[\begin{pmatrix}
    0 & -C\\
    E^{\operatorname{T}} & Q
    \end{pmatrix}_{\Surviving \Extinct}
    \begin{pmatrix}
         N_{\Extinct} \\
          R_{\Extinct}
    \end{pmatrix}-\begin{pmatrix}
    0 & -C\\
    E^{\operatorname{T}} & Q
    \end{pmatrix}_{\Surviving 0}
    \begin{pmatrix}
         N_{0} \\
         0
    \end{pmatrix}\right]\\
    &=\begin{pmatrix}
      B^{-1} & B^{-1} C_{\Surviving \Surviving} Q_{\Surviving \Surviving}^{-1}\\
   Q^{-1}_{\Surviving \Surviving} E^{\operatorname{T}}_{\Surviving \Surviving} B^{-1}&  (I-P)Q^{-1} _{\Surviving \Surviving}
    \end{pmatrix}
    \begin{pmatrix}
     -C_{\Surviving \Extinct}R_\Extinct\\
     E^T_{\Surviving\Extinct}N_\Extinct+Q_{\Surviving \Extinct}R_\Extinct-E^T_{\Surviving0}N_0
    \end{pmatrix}
\end{align}
Expanding, we have 
\begin{align}
         \delta N_{\Surviving}& = -B^{-1} C_{\Surviving \Surviving} Q_{\Surviving \Surviving}^{-1}E_{\Surviving 0}^{\operatorname{T}} N_0-B^{-1}C_{\Surviving \Extinct}R_{\Extinct}+B^{-1}C_{\Surviving \Surviving}Q_{\Surviving \Surviving}^{-1}(E^{\operatorname{T}}_{\Surviving \Extinct}N_{\Extinct}+Q_{\Surviving \Extinct}R_{\Extinct})\\
         \delta R_{\Surviving} &= -(I-P)Q^{-1}_{\Surviving \Surviving}  E_{\Surviving 0}^{\operatorname{T}} N_0 -Q^{-1}_{\Surviving \Surviving} E^{\operatorname{T}}_{\Surviving \Surviving} B^{-1}C_{\Surviving \Extinct}R_{\Extinct}  + (I-P)Q^{-1}_{\Surviving \Surviving} (E^{\operatorname{T}}_{\Surviving \Extinct}N_{\Extinct}+Q_{\Surviving \Extinct}R_{\Extinct})
         \label{eq: MCRM response}
\end{align}

\section{Linear response of nonlinear consumer resource model}
\label{app:nonlinear-model}
We can apply the linearization framework to specially non-linear consumer resource models, similarly to how we map general nonlinear ecological models to generalized Lotka-Volterra models. 

Consider general consumer resource models of the form
\begin{align}
    \frac{dN_i}{dt}&=f^N_i(N_i)g^N_i(\vec{N},\vec{R},\vec{m}) \label{eq:generalCRMNonlinN}\\
    \frac{dR_\alpha}{dt}&=f^R_\alpha(R_\alpha)g^R_\alpha(\vec{N}, \vec{R},\vec{k})
    \label{eq:generalCRMNonlinR}
\end{align}

In such general models, the per capita growth rate functions $g_i^N(\vec{N},\vec{R}, \vec{m})$ and $g^R_\alpha(\vec{N},\vec{R}, \vec{k})$ of consumer species $i$ and resources $\alpha$ respectively are nonlinear functions of consumer abundances $N_i$ and/or resource abundances $R_\alpha$. Note that even when $g^N_i$ and/or $g^R_\alpha$ are linear, the full dynamics are \emph{always nonlinear} because of the presence of the function $f^N_i(N_i)$ which is usually simply $f^R_i(N_i) = N_i$ (Eqs. (\ref{eq:generalCRMNonlinN}) and (\ref{eq:generalCRMNonlinR})). Similarly, $f^R_\alpha(R_\alpha)$ is typically---though not always---$R_\alpha$. In the remaining appendices, for brevity we might sometimes refer to such models as ``nonlinear consumer-resource models'': we mean here that the per-capita growth rate functions $g^{N}_i$ and $g^{R}_\alpha$ are nonlinear in $N_i$ and/or $R_\alpha$ (Eq. \eqref{eq:generalCRMNonlinN} and  \eqref{eq:generalCRMNonlinR})

The steady state condition for surviving species is 
\begin{align}
  0&=g^N_i(\vec{N},\vec{R},\vec{m})\\
  0&=g^R_\alpha(\vec{N}, \vec{R},\vec{k})
\end{align}
The new steady state condition is 
\begin{align}
  0&=g^N_i(\vec{N}+\delta \vec{N},\vec{R}+\delta \vec{R},\vec{m}+\delta \vec{m})\\
  0&=g^R_\alpha(\vec{N}+\delta \vec{N},\vec{R}+\delta \vec{R},\vec{k}+\delta \vec{k})
\end{align}
The linear expansion leads to 
\begin{align}
\left. \sum_j \frac{\partial g^N_i(\vec{N},\vec{R},\vec{m})}{\partial N_j}\right|_{\vec{N}=\vec{N}_{\Surviving \Surviving}}\delta N_j +\left. \sum_\alpha\frac{\partial g^N_i(\vec{N},\vec{R},\vec{m})}{\partial R_\alpha}\right|_{\vec{R}=\vec{R}_{\Surviving \Surviving}}\delta R_\alpha+ \sum_j \frac{\partial g^N_i(\vec{X}_{\Surviving \Surviving},\vec{m}_i)}{\partial{m_j}}\delta m_j&=0.\\
\left. \sum_j \frac{\partial g^R_\alpha(\vec{N},\vec{R},\vec{m})}{\partial N_j}\right|_{\vec{N}=\vec{N}_{\Surviving \Surviving}}\delta N_j +\left. \sum_\alpha\frac{\partial g^R_\alpha(\vec{N},\vec{R},\vec{m})}{\partial R_\beta}\right|_{\vec{R}=\vec{R}_{\Surviving \Surviving}}\delta R_\beta+ \sum_j \frac{\partial g^R_\alpha(\vec{X}_{\Surviving \Surviving},\vec{m}_i)}{\partial{ k_\beta}}\delta k_\beta&=0.
\end{align}
Defining each effective interaction matrix,
\begin{align}
&\left.  \frac{\partial g^N_i(\vec{N},\vec{R},\vec{m})}{\partial N_j}\right|_{\vec{N}=\vec{N}_{\Surviving \Surviving}}=\mathcal{Q}^N_{ij}, \quad \left. \frac{\partial g^N_i(\vec{N},\vec{R},\vec{m})}{\partial R_\alpha}\right|_{\vec{R}=\vec{R}_{\Surviving \Surviving}}=\mathcal{C}_{i\alpha}, \quad \frac{\partial g^N_i(\vec{X}_{\Surviving \Surviving},\vec{m}_i)}{\partial{m_j}}=\mathcal{B}^N_{ij}.\\
&\left.  \frac{\partial g^R_\alpha(\vec{N},\vec{R},\vec{m})}{\partial N_j}\right|_{\vec{N}=\vec{N}_{\Surviving \Surviving}}=-\mathcal{E}_{\alpha j},\quad \left. \frac{\partial g^R_\alpha(\vec{N},\vec{R},\vec{m})}{\partial R_\beta}\right|_{\vec{R}=\vec{R}_{\Surviving \Surviving}}=-\mathcal{Q}^R_{\alpha \beta },\quad \frac{\partial g^R_\alpha(\vec{X}_{\Surviving \Surviving},\vec{m}_i)}{\partial{ k_\beta}}\delta k_\beta=\mathcal{B}^R_{\alpha \beta}.
\end{align}
In matrix notation the linear equation becomes
\begin{align}
\mathcal{Q}^N\delta \vec{N} +\mathcal{C}\delta \vec{R}+\mathcal{B}^N\delta \vec{m}&=0.\\
-\mathcal{E}\delta \vec{N} -\mathcal{Q}^R\delta \vec{R}+ \mathcal{B}^R\delta \vec{k}&=0.
\end{align}

\begin{align}
 \begin{pmatrix}
    Q^N & \mathcal{C}\\
    -\mathcal{E} & Q^R
     \end{pmatrix}
\begin{pmatrix}
        \delta \vec{N}\\\delta \vec{R}
    \end{pmatrix}
    =
    \begin{pmatrix}
    \mathcal{B}^N & 0\\
    0 & \mathcal{B^R}
     \end{pmatrix}
    \begin{pmatrix}
        \delta \vec{m}\\\delta \vec{ k}
    \end{pmatrix}
\end{align}
\begin{align}
 \begin{pmatrix}
    (\mathcal{B}^N)^{-1} & 0\\
    0 & (\mathcal{B^R})^{-1}
     \end{pmatrix}
 \begin{pmatrix}
    Q^N & \mathcal{C}\\
    -\mathcal{E} & Q^R
     \end{pmatrix}
\begin{pmatrix}
        \delta \vec{N}\\\delta \vec{R}
    \end{pmatrix}
    =
    \begin{pmatrix}
        \delta \vec{m}\\\delta \vec{ k}
    \end{pmatrix}
\end{align}

\begin{figure*}[!hb]
\centering
\includegraphics[width=1\textwidth]{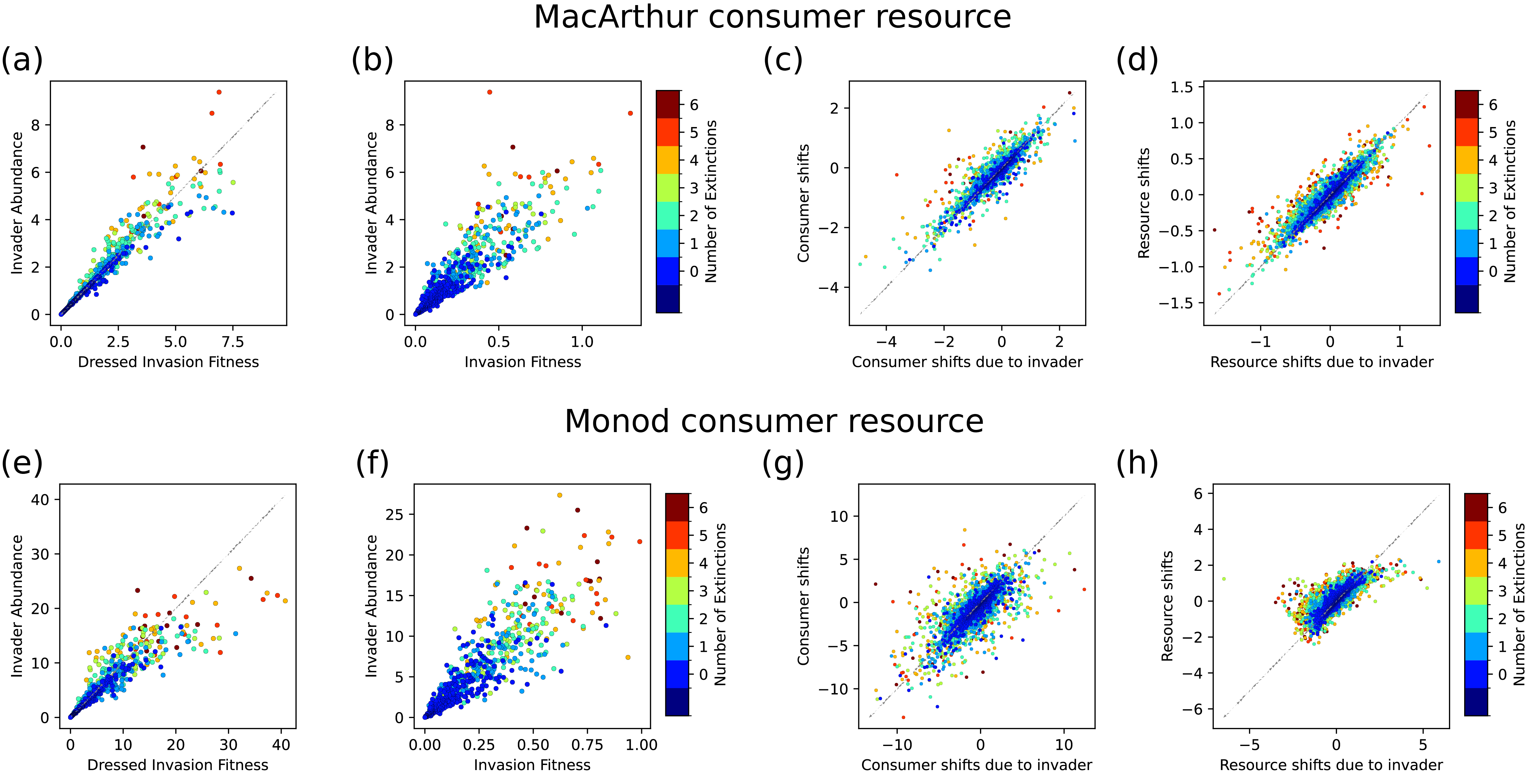}
\caption{
\textbf{ Dressed invasion fitness and extinction-induced feedbacks govern invasion outcomes in additional consumer-resource models.}  
Simulations of (a--d) the MacArthur Consumer Resource Model (MCRM) and (e--h) a Monod growth consumer-resource model show qualitatively similar results as the Microbial Consumer Resource Model (MiCRM) presented in the main text (Fig.\ref{fig:extinction effect}).
}
\label{fig:SI shift and dressed invasion fitness}
\end{figure*}

It can be solved with block matrix inversion
\begin{align}
    \begin{pmatrix}
        \delta \vec{N}\\\delta \vec{R}
    \end{pmatrix}=
    \begin{pmatrix}
     (\mathcal{B}^N)^{-1} \mathcal{Q}^N & - (\mathcal{B}^N)^{-1} \mathcal{C}\\
    (\mathcal{B^R})^{-1}\mathcal{E} & (\mathcal{B^R})^{-1}\mathcal{Q}^R
     \end{pmatrix}^{-1} \begin{pmatrix}
        \delta \vec{m}\\\delta \vec{ k}
    \end{pmatrix}
    \label{eq:SI block general}
\end{align}

For the generalized consumer resource model discussed previously, 
$(\mathcal{B}^N)^{-1}=-I$, $(\mathcal{B}^R)^{-1}=I$, $ \mathcal{Q}^N =0$,   $ \mathcal{Q}^R =Q$, $ \mathcal{C}=C$,  $ \mathcal{E}=E$.

In general, if $ \mathcal{Q}^N =0$, we can retrieve the result for nonlinear consumer resource model by substitution
\begin{align}
    C&\to -(\mathcal{B}^N)^{-1} \mathcal{C},\\
E&\to (\mathcal{B}^R)^{-1} \mathcal{E},\\
Q^R& \to   (\mathcal{B^R})^{-1}\mathcal{Q}^R.
\end{align}

\section{Linear reponse in Monod CRM with type II functional response}
\label{app: Monod derivation}
This section explicitly demonstrates the applicability of our framework to ecological models of consumer-resource dynamics where the effective interactions depend on community state as described in last section. One generic way this can occur is when the per capita growth rate functions $g_i^N(\vec{N},\vec{R}, \vec{m})$ of consumer species $i$ are nonlinear functions of consumer abundances $N_i$ and/or resource abundances $R_\alpha$ (see Eqs. (\ref{eq:generalCRMNonlinN}) and (\ref{eq:generalCRMNonlinR}) for definition). 

A paradigmatic case of such consumer-resource models is an extension of the MacArthur consumer-resource model (where $g_i$ is a linear function, even though the full dynamics are nonlinear). In this extension, consumer species dynamics $\dot{N_i}$ are governed by Monod growth rate functions (Type II functional responses) instead of linear functions of $R_\alpha$, as

\begin{align}
    \frac{dN_i}{dt} &= N_i\left(\sum_{\alpha} \frac{C_{i\alpha} R_\alpha}{h + R_\alpha} -m_i\right),\\ \frac{dR_{\alpha}}{dt} &=R_{\alpha} \left(K_\alpha-R_\alpha-\sum_j\frac{E_{j\alpha} N_j}{h + R_\alpha} \right).
\end{align}

The steady state condition is
\begin{align}
       0 &=g^N_i(\vec{N},\vec{R},\vec{m})=\sum_{\alpha} \frac{C_{i\alpha} R_\alpha}{h + R_\alpha} -m_i,\\ 0 &=g^R_\alpha(\vec{N},\vec{R},\vec{m})=K_\alpha-R_\alpha-\sum_j\frac{E_{j\alpha} N_j}{h + R_\alpha} .
\end{align}
\textbf{Effective interactions. }The main difference between Type I and Type II responses is that the consumer preferences $C_{i \alpha}$, impacts $E_{i \alpha}$, and $Q_{\alpha\beta}$ must now be replaced by effective quantities $\mathcal{C}_{i\alpha}$, $\mathcal{E}_{\alpha j}$, and $Q_{\alpha \beta}^{(R)}$ that have to be calculated from the full non-linear dynamics. These effective quantities are precisely what we call \textit{effective interactions}, since they represent how consumer species interact with resources and vice versa. Note also that $Q_{\alpha \beta}^{(R)}$ captures the effective interactions between resources. More generic models might similarly have effective interactions between species, say $B_{i j}^{(N)}$, but here $B=0$. One therefore has

\begin{align}
   \mathcal{C}_{i\alpha}= \quad \left. \frac{\partial g^N_i(\vec{N},\vec{R},\vec{m})}{\partial R_\alpha}\right|_{\vec{R}=\vec{R}_{\Surviving \Surviving}}=  \frac{C_{i\alpha} h}{(h + R_\alpha)^2},
\end{align}
 
\begin{align}
   \mathcal{Q}^R_{\alpha\beta}= -\left. \frac{\partial g^R_\alpha(\vec{N},\vec{R},\vec{m})}{\partial R_\beta}\right|_{\vec{R}=\vec{R}_{\Surviving \Surviving}}=\delta_{\alpha\beta}-\delta_{\alpha\beta}\sum_j\frac{E_{j\alpha} N_j}{(h + R_\alpha)^2},
\end{align}
and
\begin{align}
  \mathcal{E}_{\alpha j}= - \left.  \frac{\partial g^R_\alpha(\vec{N},\vec{R},\vec{m})}{\partial N_j}\right|_{\vec{N}=\vec{N}_{\Surviving \Surviving}}=\frac{E_{j\alpha} }{h + R_\alpha} .
\end{align}
Using these effective interactions, we can subsequently solve for the invader abundance in the same way as performed with the MCRM and MiCRM models. The remaining challenge lies in addressing the nonlinear invasion fitness:
\begin{align}
    f_0= \sum_{\alpha} \frac{C_{0\alpha} R_\alpha^{\text{old}}}{h + R_\alpha}-m
\end{align}
For convenience, in this model, this nonlinearity can be absorbed in the previous expression of invasion fitness simply by $C_{0\alpha}\to \frac{C_{0\alpha}}{h+R_\alpha}$.
 In summary, there are 4 replacements to normal MCRM solutions, including 
\begin{align}
    C_{i\alpha}&\to \frac{C_{i\alpha} h}{(h + R_\alpha)^2},\\
    E_{j\alpha}&\to \frac{E_{j\alpha} }{(h + R_\alpha)} ,\\
   Q_{\alpha\beta}&\to
\delta_{\alpha\beta}-\delta_{\alpha\beta}\sum_j\frac{E_{j\alpha} N_j}{(h + R_\alpha)^2}, \\
C_{0\alpha}&\to \frac{C_{0\alpha}}{h+R_\alpha} \text{, \quad for invasion fitness only},
\end{align}\\
Since all effective interaction matrices are proportional to $1/(h + R_{\alpha})$, changes in this quantity after invasion serve as a good measure of the model’s nonlinearity. In Fig.~\ref{fig:error versus C shift}, we demonstrate that the mean relative error of the framework increases linearly with the mean relative shift of this value.

\section{Linear response in the microbial consumer resource model}
\label{app: derive response MiCRM}
The Microbial consumer resource model (MiCRM) explicitly incorporates cross-feeding by modeling how metabolic byproducts from one species become resources for others. It describes the dynamics of microbial abundances $N_i$ and resource concentrations $R_\alpha$,
\begin{align}
     \frac{dN_i}{dt}&=N_i(\sum_\alpha (1-l_\alpha)w_\alpha c_{i\alpha}R_\alpha-m_i)\\
    \frac{dR_\alpha}{dt}&=K_\alpha -\omega R_\alpha - \sum_j c_{j\alpha}N_j R_\alpha +\sum_{j,\beta}l_\beta \frac{w_\alpha}{w_\beta}D_{\alpha \beta}c_{j\beta}R_\beta N_j,
\end{align}
where $l_{\alpha}$ is the fraction of consumed resource secreted as byproducts, $D_{\alpha \beta}$ is the cross-feeding matrix describing conversion between resources, $w_\alpha$ are resource weights reflecting energetic values of resources, $c_{i\alpha}$ are resource preferences, and environmental factors are contained in resource supply $K_\alpha$ and decay $\omega$. The steady state of this model, specifically the resource part, is not linear, thus requiring us to first linearize the model.

Defining effective depletion rate of resource $\alpha$ by species $i$:
\begin{align}
    \kappa_{i\alpha}^\text{eff}=-c_{i\alpha}R_\alpha+\sum_{\beta}l_\beta \frac{w_\alpha}{w_\beta}D_{\alpha \beta}c_{i\beta}R_\beta
\end{align}
The steady state condition can be written as 
\begin{align}
    0&=\sum_\alpha (1-l_\alpha)w_\alpha c_{i\alpha}R_\alpha-m_i\\
    0&=\frac{K_\alpha}{\omega} -  R_\alpha + \sum_j \frac{\kappa_{i\alpha}^\text{eff}}{\omega}N_j
\end{align}
Applying the formulation in the last section to arrive at the effective interactions in this Microbial consumer resource model, we arrive at the following effective interactions by following the same steps as in the last two appendices:
\begin{align}
    C_{i\alpha}&\to(1-l_\alpha)w_\alpha c_{i\alpha}\\
    E_{i\alpha}&\to- \kappa_{i\alpha}^\text{eff}/\omega\\
    Q_{\alpha \beta}&\to-\frac{\partial }{\partial R_{\beta}}(\frac{K_\alpha}{\omega} -  R_\alpha + \sum_j \frac{\kappa_{i\alpha}^\text{eff}}{\omega}N_j)\\
    &=\delta_{\alpha \beta}(1 +\frac{1}{\omega}\sum_j c_{j\alpha}N_j)-l_\beta \frac{w_\alpha}{w_\beta}D_{\alpha \beta}\sum_{j}c_{j\beta} N_j
\end{align}

\section{Algorithm to solve the self-consistency equations}
Here we present a simple algorithm to solve the self-consistency equations iteratively. Algorithm~\ref{alg:general_perturbation} outlines the basic logic and implementation steps. We assess convergence by monitoring the extinction boolean indicators that determine the extinction and survival status of each species. The iteration terminates when these indicators remain unchanged following an update, indicating that the extinction and survival states have stabilized. If convergence is not reached, we impose a maximum of 50 iterations, after which we accept the last extinction boolean prediction as the final result. While Algorithm~\ref{alg:general_perturbation} performs reasonably well on its own, we incorporate several numerical improvements to enhance stability and better accommodate complexities arising in simulation results and experimental data:

First, when calculating the abundance shifts of resident species, we introduce a momentum hyperparameter at step 6. We chose its value to be $0.1$, meaning the new iteration contributes $10\%$ weight while the accumulated value carries $90\%$ weight. This parameter incorporates a weighted contribution from the abundance shift of the previous iteration into the current update, smoothing fluctuations and accelerating convergence. Such momentum-based updates are a well-known technique in numerical optimization to improve stability. However, to truly reflect the self-consistency equations, we use the final extinction boolean to recompute the abundances of both invaders and surviving species, so the memory from previous iterations is only retained in this boolean state.

Second, numerical uncertainties in abundance data before invasion mean that extinct species may not reach exact zero abundances in practice. To address this, at step 7 we implement a survival threshold, a small cutoff value ranging from $10^{-3}$ to $10^{-7}$ depending on the magnitude of species abundances in the particular model. Species with predicted abundances below this threshold are treated as extinct, instead of below exactly zero as described in step 7. This step effectively handles numerical noise and improves extinction boolean predictions.

Third, for models with nonlinear invasion fitness terms, such as Monod consumer resource model, we provide an additional input argument representing the effective consumer preference matrix used to compute nonlinear invasion fitness. This matrix is distinct from the ordinary interaction matrix 
$A$ and extend to complex models (see Appendix~\ref{app: Monod derivation} for the detail example for Monod model).

Together, these modifications improve the robustness and applicability of the iterative algorithm across diverse systems, allowing accurate prediction of species abundances and extinction patterns following invasion.

The full code implementation underlying all results in this paper is publicly available at \url{https://github.com/Emergent-Behaviors-in-Biology/invasion-theory}

\begin{algorithm}[H]
\caption{Iterative solver for self-consistency equations}
\label{alg:general_perturbation}
\begin{algorithmic}[1]
\State \textbf{Input:} Pre-invasion system interaction matrix $A$, species abundances $\vec{X}$, growth rates $\vec{r}$, invader interaction matrices $A_{II}$, $A_{IS}$, $A_{SI}$, and invader growth rates $u_I$, maximum number of iterations $\texttt{max\_num\_iters}$.
\State \textbf{Initialize:} Set extinction boolean $Ebool$ to \texttt{False} for all species (i.e., assume no species are extinct)
\For{$i = 1$ to $\texttt{max\_num\_iters}$}
    \State According to $Ebool$, extract submatrices and vectors for surviving, extinct, and invader species as described in Eq.~\eqref{eq: 9 blocks}
    \State Solve invader abundances using Eq.~\eqref{eq:invader abundance}
    \State Compute abundance shifts for the surviving community using Eq.~\eqref{eq:community shift}
    \State Update extinction boolean $Ebool$: 
        \begin{itemize}
            \item Set $Ebool = \texttt{True}$ for species with negative abundance
            \item Set $Ebool = \texttt{False}$ for species with positive invasion fitness
        \end{itemize}
    \If{\textbf{convergence condition met}}
        \State Break loop
    \EndIf
\EndFor
\State \Return Final predicted invader abundances, community shift, and extinction boolean
\end{algorithmic}
\end{algorithm}

\section{Error analysis}

\begin{figure*}[ht]
\centering
\includegraphics[width=0.8\textwidth]{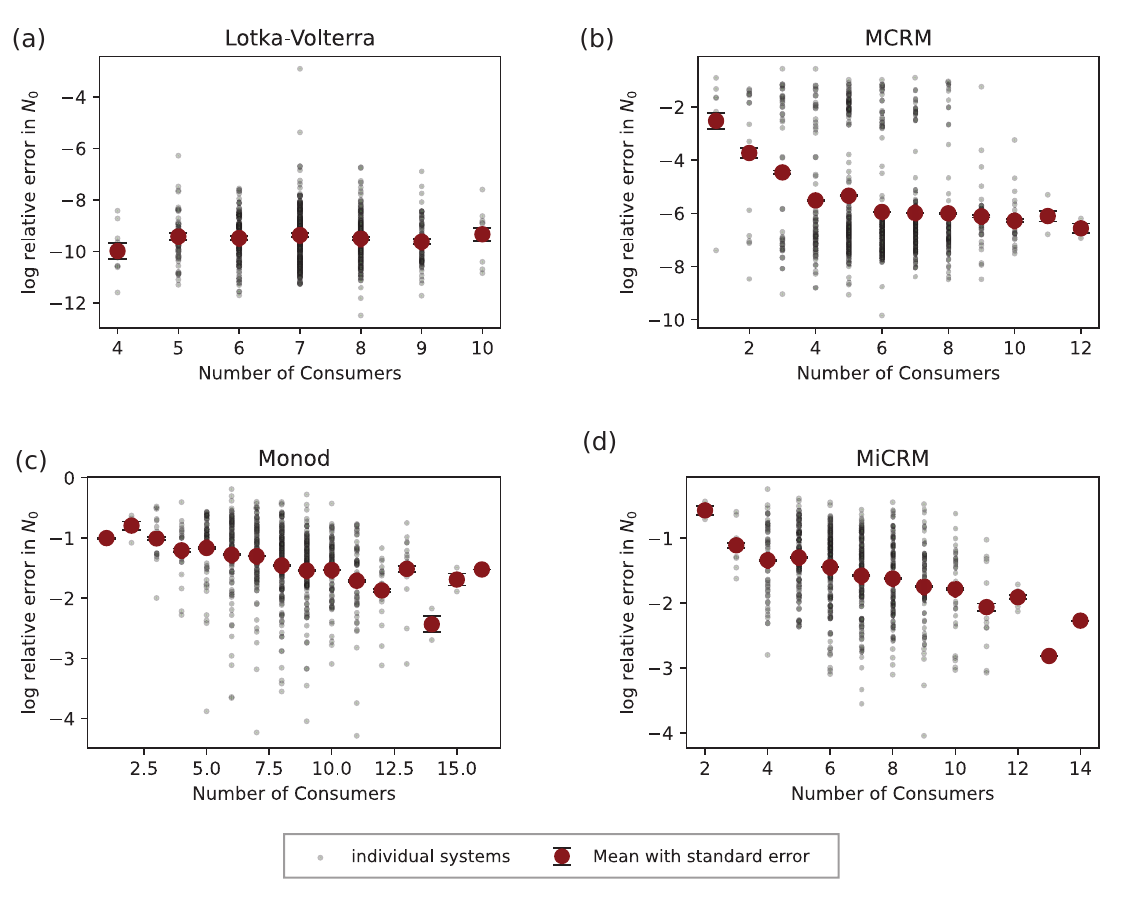}
\caption{
\textbf{ Error decreases with system size}  With the same simulation data used in Fig.\ref{fig:all models agree}, we grouped predictions into different number of consumers after the invasions. For (a) Lokta-Volterra and (b) MacArthur consumer resource models, the decreasing trend saturate when the system is large enough because the self-consistency equation is exact and errors are only coming from the algorithm to deduce identity of extinct species, which also improves with system size. For (c) consumer resource models with Monod functional response and (d) Microbial consumer resource models, the error shows consistent decreasing trend because of the linear-response approximation to non-linear models.}
\label{fig:error plot}
\end{figure*}

The errors observed in the predictions compared to simulation results in Fig.~\ref{fig:all models agree} originate from multiple sources. Broadly, the error of the framework prediction arise from two components: (1) the self-consistency equations themselves and (2) the algorithm used to solve them. The primary source of error may vary depending on the particular model. Theoretically, the self-consistency equations are exact for models with linear steady-state conditions, including GLV Eq.~\eqref{eq:invader abundance},
Eq.~\eqref{eq:community shift} and MCRM Eq.~ \eqref{eq: MCRM invader abundance} to \eqref{eq: MCRM response} . To demonstrate this, Fig.~\ref{fig:MCRM_work_with_info} presents the results of solving the equations for the MCRM model using the same simulations as in Fig.~\ref{fig:all models agree}(e)--(g), but with the extinction and survival states of each species after invasion provided explicitly rather than inferred through our iterative algorithm. This is done to disambiguate errors in our algorithm and approximations in our self-consistency equations. Notably, all visible errors in Fig.~\ref{fig:all models agree}(e)--(g) disappear once the error introduced by the algorithm is removed (except numerical errors coming from ODE simulations themselves, between a relative  $10^{-9}$ to $10^{-2}$. Since our iterative algorithm explores the boolean space of extinction and survival states locally, beginning from an initial condition where all resident species are assumed to survive, its performance improves when the true states do not deviate substantially from this starting point. Consequently, the mean relative error of invader abundance in Fig.~\ref{fig:error plot}(b) declines with number of consumers after the invasion, despite the error distribution remaining bimodal, with peaks centered around $10^{-7}$ (likely due to numerical simulation uncertainty) and $10^{-2}$ (attributable to the algorithm) respectively. A similar trend is observed in the Lotka-Volterra model shown in Fig.~\ref{fig:error plot}(a), although only two instances of substantial error occur, both at small consumer numbers of 4 and 5. 

\begin{figure*}
\centering
\includegraphics[width=0.4\textwidth]{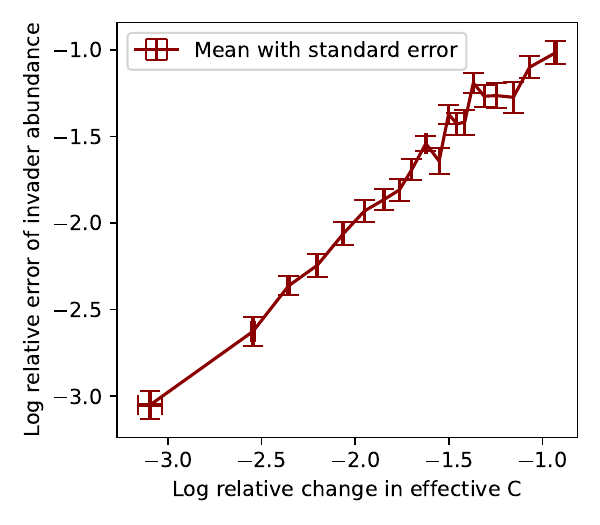}
\caption{
\textbf{ Error increases with degree of nonlinearity in Monod consumer resource model}  The effective interaction in Monod consumer resource model are all  proportional to $1/(h+R)$, The mean relative shift in this quantity serves as a direct measure of nonlinearity, quantifying the extent to which our linear-response approximation breaks down as nonlinearity increases. }
\label{fig:error versus C shift}
\end{figure*}

\begin{figure*}
\centering
\includegraphics[width=\textwidth]{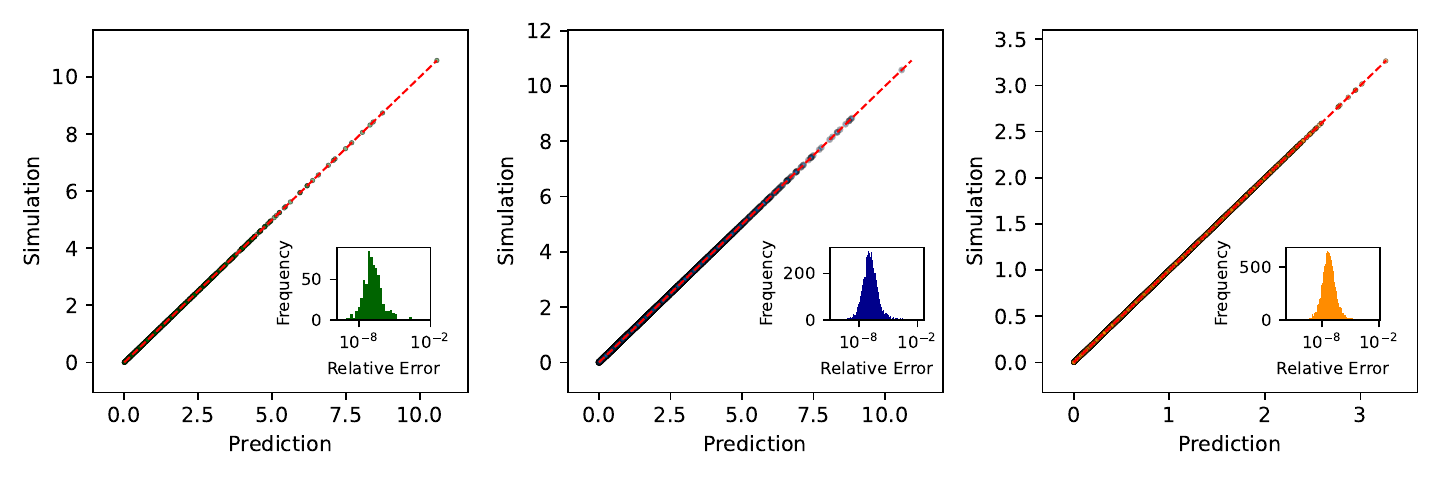}
\caption{
\textbf{ Perfect prediction of MCRM when correct information of extinction is given.} We demonstrate that the error distribution in Fig.~\ref{fig:all models agree}(e)--(g) only has a larger second peak because of wrong boolean prediction of extinctions of species. When this information is given, the larger peak disappear and left with the smaller peak that only comes from numerical uncertainty of simulations.}
\label{fig:MCRM_work_with_info}
\end{figure*}

For nonlinear models such as MiCRM and Monod, the primary source of error likely arises from the self-consistency equations using a linear response approximation. This approximation assumes that the effective interactions among surviving species do not change significantly after invasion as described from Eq.\eqref{eq:new steady state} to \eqref{eq: simplified general perturbation}. However, since these effective interactions depend on species or resource abundances, the resulting solutions from the self-consistency equations deviate from the actual invasion outcomes. This effect is clearest to test in the Monod model. As demonstrated in Fig.~\ref{fig:error versus C shift}, the mean of the logarithm of the relative error increases nearly linearly with the mean of the logarithm of relative change in the effective interactions, confirming that deviations in these interactions contribute directly to the error. Moreover, this linear response approximation becomes more accurate in more species-diverse ecosystems, as the relative shift in species abundance caused by a single invader diminishes with increasing species number. This consistent trend is also evident in Fig.~\ref{fig:error plot} (c)--(d), where the logarithm of the relative error decreases as the species diversity (number of surviving consumer species) increases.

\section{Numerical details for simulations}
\subsection{Lotka-Volterra Model}
\label{app: LV numerical}
In Fig.~\ref{fig:all models agree}(b)-(c), and Fig.~\ref{fig:extinction effect}(c)-(e), we simulated an ensemble of 640 Lotka-Volterra models. We assembled each system's pre-invasion community with $n=10$ species, which resulted in an average of $8$ surviving species. The interaction matrices followed statistics 
\begin{align}
    \langle A_{ij}\rangle =\mu_A/n + \delta_{ij}, \langle A_{ij}A_{kl}\rangle =\delta_{ij}\delta_{kl}\frac{\sigma_A^2}{n}+\delta_{il}\delta_{jk}\frac{\rho\sigma_A^2}{n}.
\end{align}
And the environmental parameters similarly were independent and had mean $\mu_r/n$ and variance $\sigma_r^2/n$. In this figure we chose $\mu_A=2$, $\sigma_A=0.2$, $\rho=1$ for a symmetric model, and $\mu_r=2$, $\sigma_r=0.1$. We chose the threshold for extinction as $10^{-5}$. Invaders were uncorrelated to the system but shared the same statistics of interaction matrix and environmental parameters. We performed simulations with numerical ODE simulator 'LSODA', with parameters atol=1e-12, rtol=1e-12, and checked convergence by ensuring the maximum derivative was less than 1e-12. 

\subsection{MacArthur consumer resource model}
\label{app: CRM numerical}
In Fig. ~\ref{fig:all models agree}(e)--(g) and Fig.~\ref{fig:SI shift and dressed invasion fitness}(a)--(d), we simulated an ensemble of 640 generalized MacArthur consumer resource models. We assembled each system's pre-invasion community with $60$ consumers and $n=30$ resources, which resulted in an average of 7 surviving consumers and 22 surviving resources. In this model, we set consumer extinction threshold at $10^{-3}$ and resource extinction threshold at $10^{-9}$. The consumer preference matrix followed statistics: 
 \begin{equation}\langle C_{i\alpha}\rangle =\frac{\mu_C}{n}, \langle C_{i\alpha}C_{j\beta}\rangle =\delta_{i\alpha}\delta_{j\beta}\frac{\sigma_C^2}{n}\end{equation}
In the figure we used $\mu_C=60$, $\sigma_C=0.5$. For demonstration purpose of the generality, the resource had a self-interaction matrix $Q$ with statistics, 
 \begin{equation}\langle Q_{ij}\rangle =\mu_Q/n+\delta_{ij}\end{equation}
  \begin{equation}\langle Q_{ij}Q_{kl}\rangle =\delta_{ij}\delta_{kl}\frac{\sigma_Q^2}{n}+\delta_{il}\delta_{jk}\frac{\rho_Q\sigma_Q^2}{n},\end{equation}
  where we used $\mu_Q= 0.1, \sigma_Q=0.1, \rho_Q=0.1$.

For environmental parameters, we had mean and variance of consumer mortality rate, $\mu_m/n$, $\sigma_m^2/n$, with $\mu_m=1$, $\sigma_m=0.1$. And mean and variance of carrying capacity were, $\mu_k/n$, $\sigma_k^2/n$, with $\mu_k=2$, $\sigma_k=0.1$. Again, the invader parameters were uncorrelated with the system and followed the same statistics. We used a constraint optimization algorithm instead of numerical ODE for simulation. The algorithm was similar to the one used in \cite{Marsland2020}.

\subsection{Microbial consumer resource model}
\label{app: MiCRM numerical}

In Fig. \ref{fig:all models agree} (m)-(o), Fig.~\ref{fig:extinction effect} (f)-(i), and Fig.~\ref{fig: evolved} (b)-(e) with legend ``Assembled'', we simulated an ensemble of 640 microbial consumer resource models (MiCRM). We assembled each system's pre-invasion community with $60$ consumers and $n=30$ resources, which resulted in an average of 8 surviving consumers, while resources were abiotic and did not go extinct. It had a cross-feeding structure where we supplied one of the resources at a high rate $k=1000$, while others were by-products. 
Similar to before, the consumer preference matrix was \begin{equation}\langle c_{i\alpha}\rangle =\frac{\mu_c}{n}, \langle c_{i\alpha}c_{j\beta}\rangle =\delta_{i\alpha}\delta_{j\beta}\frac{\sigma_c^2}{n}\end{equation}
with $\mu_c=10$, $\sigma_c=3$, the leakage rate was a constant $l=0.8$ across consumers. We chose the sparsity parameter for cross-feeding matrix to be $0.2$. We fixed extinction threshold abundance for consumers at 0.1. We performed the simulation with community simulator, which used an Expectation-Maximization algorithm that utilized a constraint optimization algorithm to find the ecological steady states.

\subsection{Eco-evolutionary dynamics of MiCRM}
\label{app: evo numerical}
In Fig. \ref{fig: evolved} (b)-(h), we performed a series of simulations for eco-evolutionary dynamics for 40 successful invasions on an ensemble of 640 microbial consumer resource models. We first initialized the communities with parameters whose statistics were identical to communities from the previous section on the microbial consumer resource model (MiCRM); we removed all extinct species at an ecological steady state. Then, at each iteration, we randomly selected a parent from the surviving consumers, and sampled a mutant as an invader with consumer preference vector $\vec{c}_{\text{mutant}}$ correlated with a randomly selected parent $\vec{c}_{\text{parent}}$ according to the equation 
\begin{align}
    \vec{c}_{\text{mutant}}=\mu_c +\rho  (\vec{c}_{\text{parent}}-\mu_c) + \sqrt{1 - \rho^2} \vec{z}_c
\end{align}
where $\rho$ is the correlation between mutant and parent traits, and $z_c$ is a random vector composed of uncorrelated Gaussian variables with zero mean and standard deviation $\sigma_c/\sqrt{n}$. For our analysis, we used $\mu_c=10$ and $\rho=0.8$. After we sampled the invader and added it to the community, we solved for the new ecological steady-state using the iterative optimization algorithm common to all MiCRM simulations. We removed extinct species if they were below a threshold abundance of $0.1$. We repeated this iterative eco-evolution until we observed 40 successful invasions in each community.

In Fig.~\ref{fig: evolved}(b), we applied hierarchical clustering to the consumer preference matrix to group consumers and resources based on their pairwise correlations. We computed Pearson correlation coefficients for rows (consumers) and columns (resources) separately, and performed clustering using the average linkage method. We reordered the matrix based on the resulting dendrograms to highlight correlation of similar consumers and resources. We visualized the reordered matrix as a heatmap with interaction strengths represented by color intensity.

In Fig.\ref{fig: evolved}(c), we calculated the average extinction probability versus consumer abundance by grouping the consumers into bins according to their pre-invasion abundance and computing the average extinction probability for consumers within each bin. The bins are grouped with approximately the same number of samples per bin. Number of bins in this plot is chosen to be 13, with around 350 samples per bin for the evolved systems, and around 460 for the assembled systems. The error bars represent the standard error for both axes, calculated for each bin.

\subsection{Consumer resource model with (Monod) Type-II functional response }
In Fig.~\ref{fig:all models agree} (i)--(k), and Fig.~\ref{fig:SI shift and dressed invasion fitness}(e)--(h), we simulated an ensemble of 640 consumer-resource models employing Monod (Type-II) functional responses. The pre-invasion communities were sampled from a regional species pool consisting of 30 consumers, with the number of resources set to $n=30$. The Monod constant $h$ was chosen as $h=3$, comparable in magnitude to the resource abundances, thereby introducing a moderate degree of nonlinearity. Consumer preferences were sampled similarly to other models, with parameters $\mu_c = 3$ and $\sigma_c = 1$. Environmental variables were held constant across species, with values $k = 3$ and $u = 1$.

In Fig.~\ref{fig:error versus C shift}, we varied $h$ to sample models with different non-linearity. We choose $10$ values of $h$ linearly from $h=3$ to $h=10$. For each value we sampled $100$ systems. Other parameters are kept the same as before. 
All simulations for this model is performed with numerical ODE simulator 'LSODA', with parameters atol=1e-11, rtol=1e-11, and we checked convergence by ensuring the maximum derivative was less than 1e-10. 

\section{Methods for Data Analysis and Visualization}
In Fig. \ref{fig:data}, we used species abundance data obtained from three different experimental works from Kuebbing et al.\cite{kuebbing2015above}, Rakowski and Cardinal \cite{rakowski2016herbivores}, and Pennekamp et al. \cite{pennekamp2018biodiversity}. In addition, we used the Lotka-Volterra interaction matrix fitted for these datasets in a previous work \cite{maynard2020predicting}.

We repeated the same set of procedures for these 3 different datasets as follows: (1) we first took the median abundance of each species in the largest ecosystem (containing all species) as the true values corresponding to new community abundance $X_S^{\text{new}}$ and invader abundance $X_0$, plotted on the y-axis; (2) then we considered the abundance of smaller ecosystems (assemblages) with exactly one fewer species than the largest system and treated the missing species as an invader to those smaller systems as $X_S^{\text{old}}$; (3) Using our framework described by Eq. \eqref{eq:4 block}-\eqref{eq:surviving-SCE}, we predicted $X_S^{\text{new}}$ and $X_0$. Since for each dataset, researchers repeated the same experiment with the same community composition (assemblages) 5 to 10 times, we had a prediction for each of these replicas. In Fig. \ref{fig:data}, we plotted the median values on x-axis, and used the 25th and 75th percentiles as error bars.

\section{Additional experimental data}
To further assess the applicability of our framework to experimental systems, we present results for three additional sets of biodiversity-ecosystem functioning (BEF) experimental data \cite{tilman2001diversity, van2010diversity, cadotte2013experimental}. We used Lotka-Volterra interaction matrices  fitted in prevoiusly work \cite{lemos2024phylogeny}, where phylogenetic relationships are assumed to shape species competition. Among the four models fitted with similar performance, we selected Model 3, which provides an intermediate level of complexity. Each dataset comprised species assemblages from distinct ecosystems. Compared to the datasets analyzed in the main text, we found two key differences. First, most assemblages lacked replicates, preventing us from estimating variability or error bars. Second, the datasets did not include exhaustive combinations of all possible species; consequently, pairs of assemblages used for comparison may differ by multiple species, with several considered as invaders or extinctions.

\begin{figure}
\centering
\includegraphics[width=0.5\textwidth]{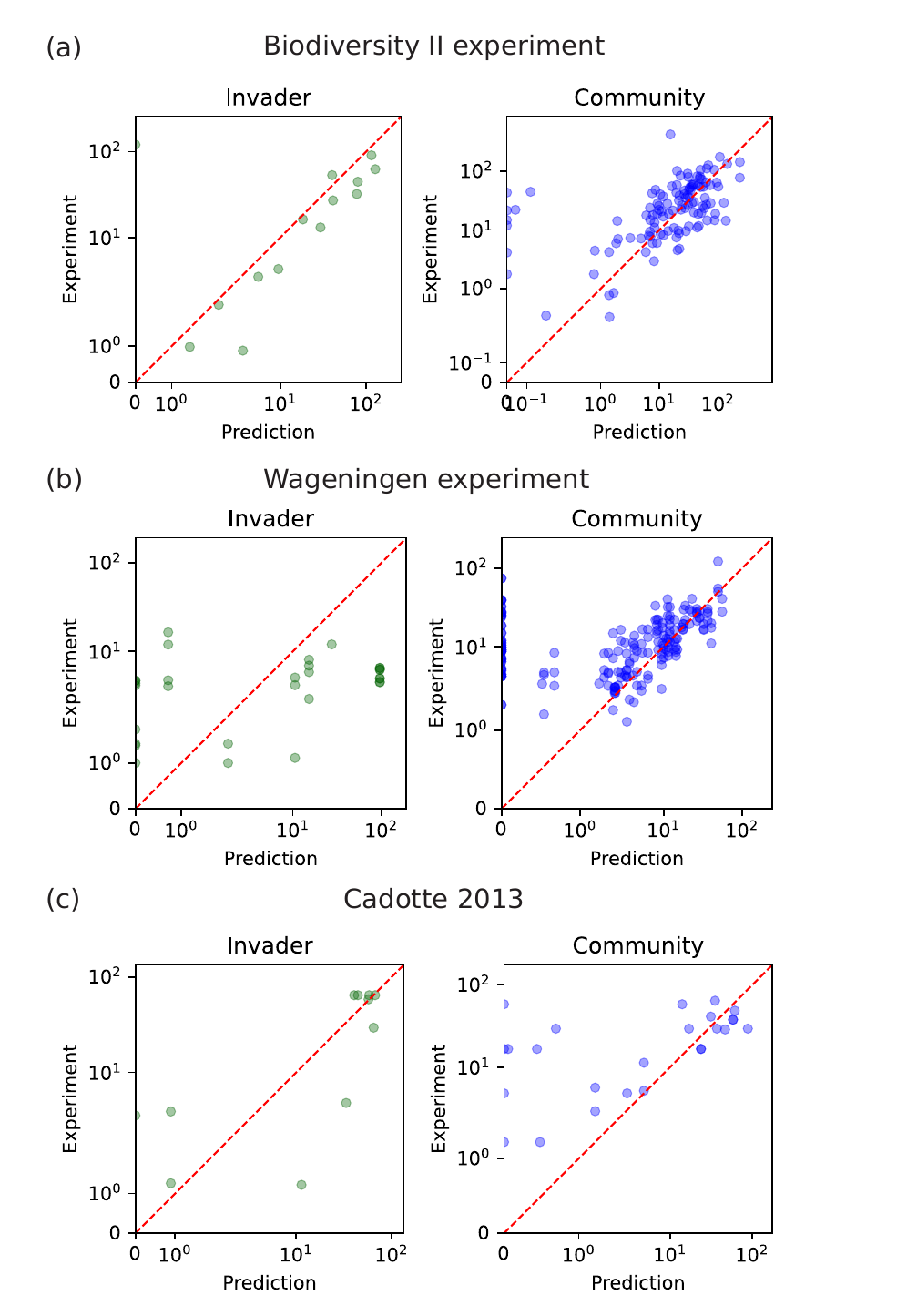}
 \caption{\textbf{Theory applicable to bio-diversity ecosystem and functioning (BEF) experiments.} Scatter plots comparing  experimentally measured abundances with theoretical predictions for three datasets of assemblages of plant ecosystems. We considered pairs of assemblages which differed from each other by the introduction of exactly one species which we call the invader, and variable numbers of extinctions events. Using our theory for the GLV model on these datasets, we predict both the invader abundance (left) and post-invasion community abundances (right). There is no replicates in general and therefore no error bars. Both axes use a sym-log scale, with linear scaling applied below the smallest non-zero data point. The datasets are all plant communities from different experiments, filtered to pairs of assemblages with high species similarity (defined as ratio of shared species in two assemblages) for visual clarity, (a) 9 years of Biodiversity II experiment \cite{tilman2001diversity} with similarity  larger than 0.85, (b) 9 years of Wageningen experiment\cite{van2010diversity} with similarity larger than 0.85  and (c) one dataset Cadotte 2013 with similarity larger than 0.6. See Fig.~\ref{fig:full extra data} for all pairs of assemblages with similarity larger than 0.5 \cite{cadotte2013experimental}.
}
\label{fig:extra data}
\end{figure}

\begin{figure}
\centering
\includegraphics[width=0.5\textwidth]{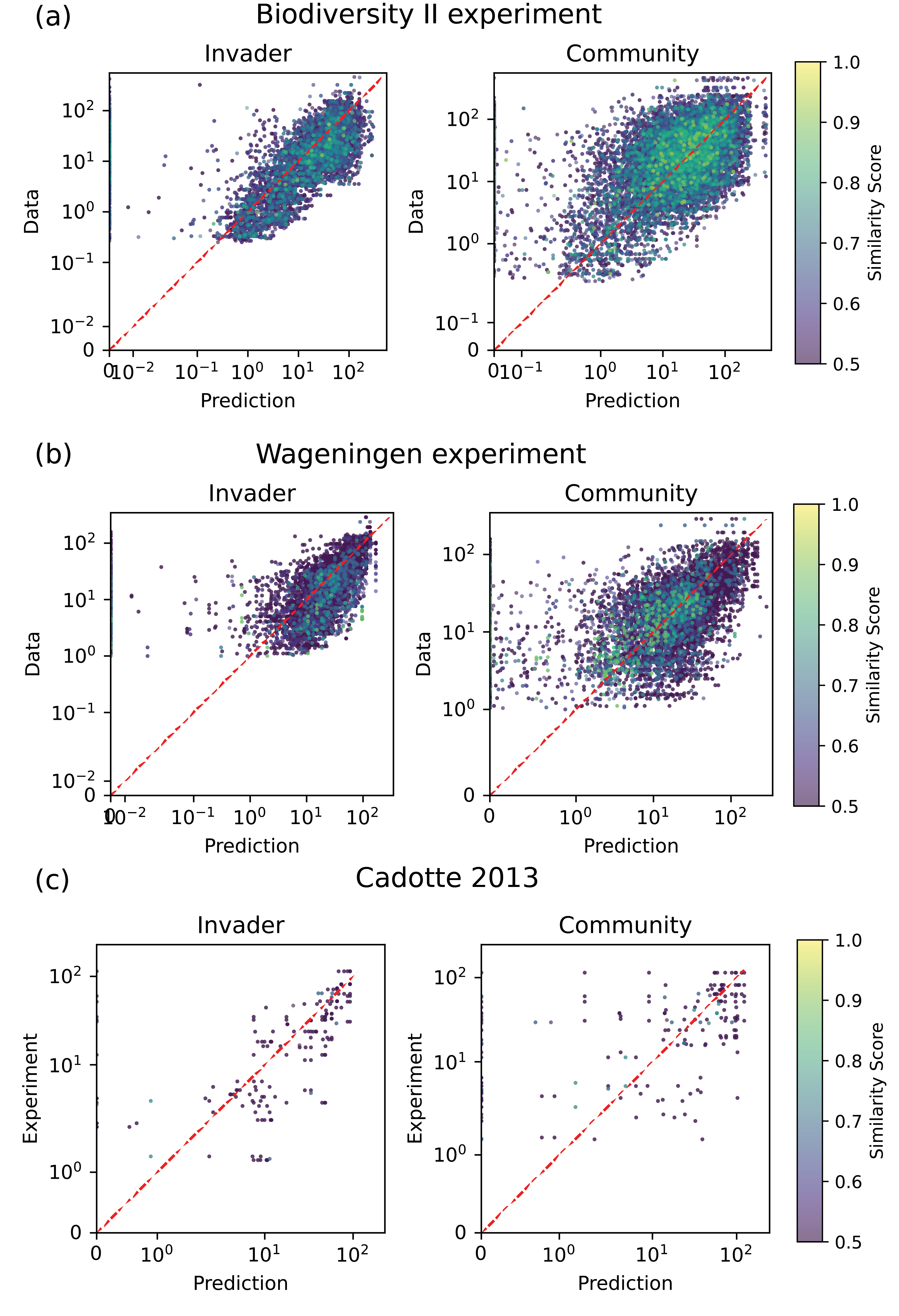}
 \caption{ \textbf{Extended version of Figure \ref{fig:extra data} with variable number of invaders and number of extinction events.} The marker color indicates the similarity of species between the pairs of assemblages, defined as the number of shared species divided by the total number of species in these assemblages. Only assemblages with more than half of shared (similarity$>0.5$)  species are included. The plots are (a) 9 years of Biodiversity II experiment \cite{tilman2001diversity} with  (b) 9 years of Wageningen experiment \cite{van2010diversity}, and (c) one dataset Cadotte 2013 \cite{cadotte2013experimental}.
}

\label{fig:full extra data}
\end{figure}

To align with the assumptions of our framework, we filtered pairs of assemblages based on ``similarity'', defined as the fraction of shared species relative to the total in either assemblage. In Fig.~\ref{fig:extra data}, to enable comparable analyses to the main text, we first limited to systems with a clearly defined single invader, then applied different similarity thresholds to show a comparable number of invasion results on the plots. For (a), covering 9 years of the Biodiversity II experiment \cite{tilman2001diversity} and 9 years of the Wageningen experiment \cite{van2010diversity}, a similarity threshold of 0.85 was applied. For (c), corresponding to the Cadotte 2013 dataset \cite{cadotte2013experimental}, we used a threshold of 0.6. Additionally, in Fig.~\ref{fig:full extra data}, we plotted all pairs of assemblages with a similarity above 0.5, coloring points by similarity value. There is a clear trend showing that the similarity value positively correlates with the prediction performance, meaning that pairs of assemblages with higher species overlap tend to yield more accurate predictions within our framework.

\begin{figure*}
\centering
\includegraphics[width=0.7\textwidth]{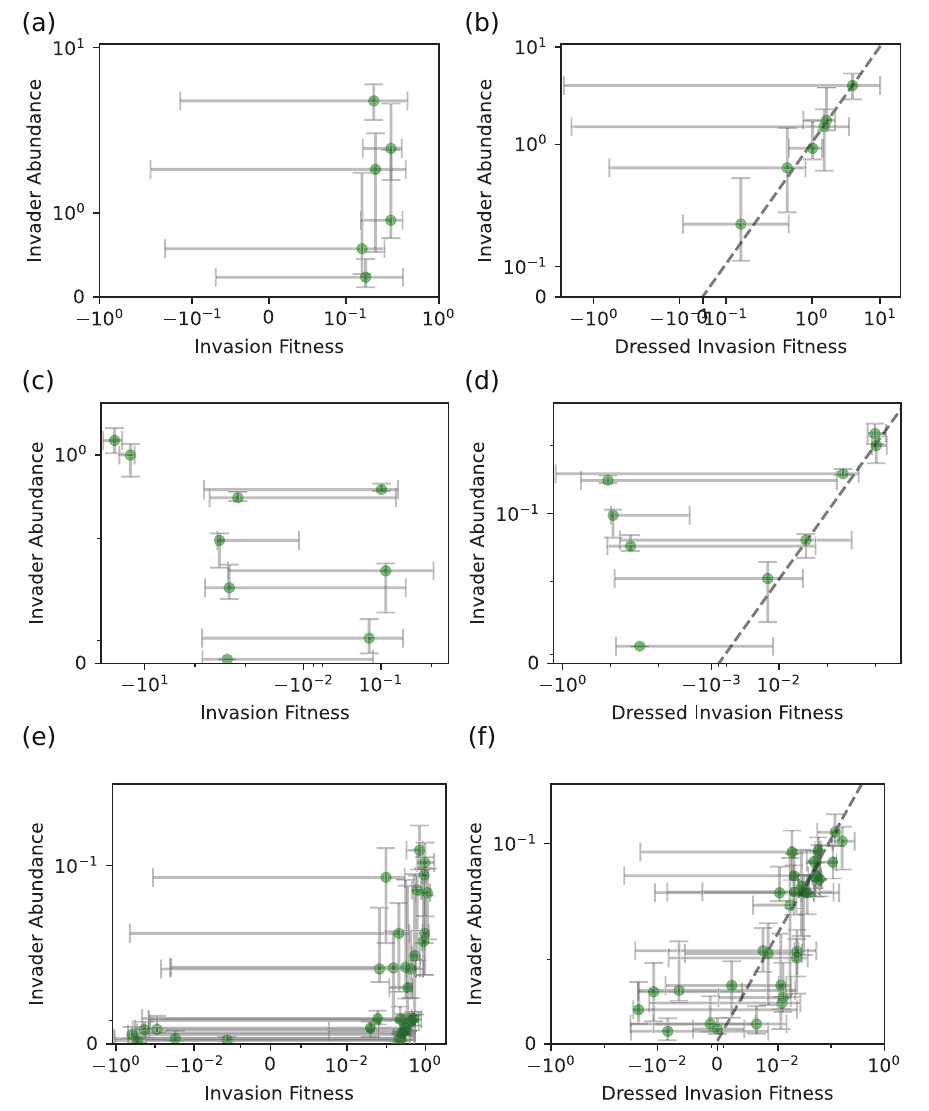}
\caption{
\textbf{ Dressed invasion fitness correlates more strongly with invader abundance in experimental data.} Scatter plots comparing  experimentally measured invader abundances with invasion fitness and dressed invasion fitness, which are both computed from the fitted LV models and the pre-invasion abundance.  Points indicate median value for measurements, with error bars indicating variability across replicates. Both axes use a sym-log scale, with linear scaling applied below the smallest non-zero data point. The datasets span (a)--(b) plant communities \cite{kuebbing2015above}, (c)--(d) herbivore–algae communities \cite{rakowski2016herbivores}, and (e)--(f) ciliated protist communities \cite{pennekamp2018biodiversity}.
}
\label{fig: data dressed invasion fitness}
\end{figure*}

\begin{figure*}
\centering
\includegraphics[width=0.7\textwidth]{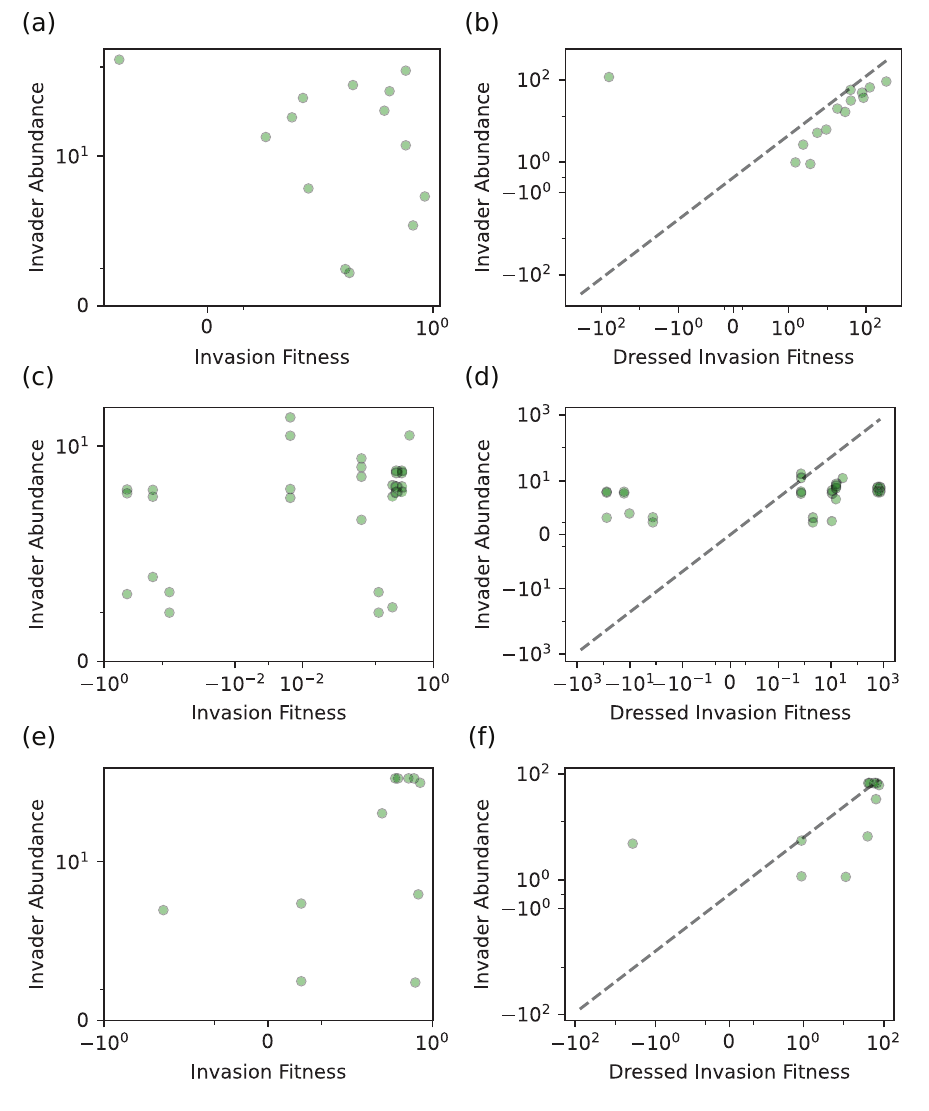}
\caption{
\textbf{Additional data of dressed invasion fitness correlates more strongly with invader abundance in experimental data.} Comparison of the relationship between invader abundance and both dressed invasion fitness and invasion fitness, shown in analogous plots to Fig.~\ref{fig: data dressed invasion fitness}, for additional plant experiments: (a)–(b) nine years of the Biodiversity II experiment \cite{tilman2001diversity} with similarity larger than 0.85, (c)--(d) nine years of the Wageningen experiment \cite{van2010diversity} with similarity larger than 0.85, and (e)--(f) one dataset from Cadotte 2013 with similarity larger than 0.6. See Fig.~\ref{fig:full extra data} for all pairs of assemblages with similarity larger than 0.5 \cite{cadotte2013experimental}.
}
\label{fig: extra data dressed invasion fitness}
\end{figure*}

\begin{figure}
\centering
\includegraphics[width=0.5\textwidth]{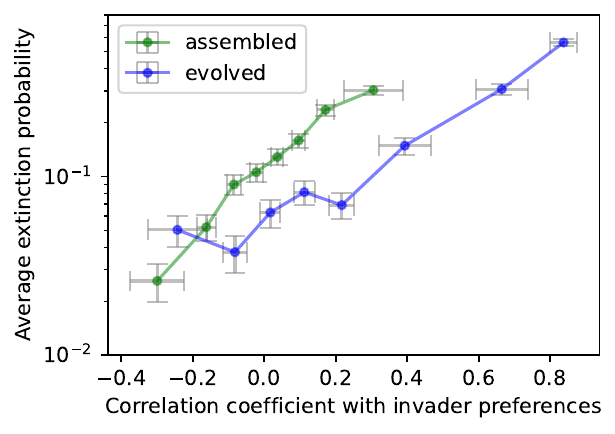}
 \caption{ \textbf{Extinction probability increases with invader-resident correlation, consistent with competition exclusion.} We used the same simulation results  from the MiCRM model in Fig. \ref{fig: evolved} to plot how the extinction probability of a resident species depends on its phenotypic similarity with an invader. Similarity is measured using correlation between consumption preferences of residents and invaders. For evolved communities, the highest correlation would be between parents (residents) and their mutants (invaders). The plot shows that resident species are indeed more likely to become extinct when their consumer preferences are more correlated with an invader, consistent with competitive exclusion \cite{sireci2023environmental, lemos2024phylogeny}. However, note that even at high correlation, extinction probability is still $\sim 60\%$. This indicates that in roughly 40\% of cases, invaders can still coexist with highly similar species as long as they are embedded in a diverse community.
}
\label{fig:extinction versus correlation}
\end{figure}

\end{document}